\documentclass[prb,twocolumn]{revtex4-1}
\usepackage{graphicx,bm,float,color,hyperref,amssymb,amsmath,physics,mymacros}
\hypersetup{colorlinks=true linkcolor={blue}}

\begin{document}

\title{High fidelity ac gate operations of a three-electron double quantum dot qubit}
\author{Clement H. Wong}
\affiliation{Department of Physics, University of Wisconsin-Madison, Madison, Wisconsin 53706, USA}

\date{\today}
\begin{abstract}
Semiconductor quantum dots in silicon are promising qubits because of long spin coherence times and their potential for scalability.  However, such qubits with complete electrical control and fidelities above the threshold for quantum error correction have not yet been achieved.  We show theoretically that the threshold fidelity can be achieved with ac gate operation of the quantum dot hybrid qubit.  Formed by three electrons in a double dot, this qubit is electrically controlled, does not require magnetic fields, and runs at GHz gate speeds.  We analyze the decoherence caused by 1/f charge noise in this qubit, find parameters that minimize the charge noise dependence in the qubit frequency, and determine the optimal working points for ac gate operations that drive the detuning and tunnel coupling. 
\end{abstract}
\maketitle

\section{introduction}

Silicon semiconductor quantum dot qubits are promising for quantum information processing because of their long spin coherence  times and their potential for scalability and integration with classical electronics.\cite{zwanenburgRMP13} 
While high fidelity spin qubits in silicon have recently been achieved, \cite{kawakamieNATN14,veldhorstmNATN14} for a practical quantum computer,
it is also desirable to have a qubit that is purely electrically controlled, that does not require magnetic fields, and possess fast gate speeds.
The quantum dot hybrid qubit~\cite{shiPRL12,kohPRL12,{shiNATC14},kimNAT14} has the potential to meet these criteria. 
This qubit, which consists of three electrons in a double dot, has a qubit frequency of $11.5$ GHz, set by the single dot singlet triplet splitting, and allows complete qubit control via the detuning--the voltage bias between the dots, and the tunnel coupling between the dots.  
Recent experiments have demonstrated $\simgt85\%$ gate fidelities for dc (direct current) operation  and $\simgt 93\%$ for resonant ac (alternating current) operation that drives the detuning.
The gate operations that limit the qubit fidelity in the these experiments involve transitions between qubit states, which we call $X$ rotations. 
%These rotations do not benefit from the fast rotation speed set by $E_{ST}$.
While ac gate operations, which drive these transition in a manner similar to  electron spin resonance, have resulted in a significant improvement in qubit fidelity,  fidelities above the threshold for quantum error correction have not yet been achieved. 

To understand the limiting factors in ac gate operations, one needs to examine decoherence for driven qubits, which is substantially modified from the case of free qubit evolution.
This subject has been studied extensively both theoretically and experimentally for superconducting qubits. \cite{yanNATC13,ithierPRB05,smirnovPRB03}
In contrast to free evolution, where typically all noise power below the precession frequency contribute to dephasing, during driven evolution, the effect of low frequency noise on the qubit dynamics is mitigated, resulting in significantly improved coherence times.
On the other hand, the driven qubit is exposed to noise at specific high frequency components in the noise spectrum.
In particular, dephasing of ac resonant gates for the hybrid quantum dot qubit is sensitive to the noise power at the \emph{qubit} frequency.

In this paper, we perform a systematic analysis and numerical optimization of ac resonant gate fidelities for the  quantum dot  hybrid qubit.  
We develop a decoherence model for the hybrid qubit that fully takes into account the $1/f$ charge noise spectrum, which is the dominant source of decoherence in double quantum dots.
Our model also takes into account nonlinear qubit dynamics that occur under strong driving conditions.
While we consider only the hybrid qubit in this paper, our model can be readily applied to other semiconductor qubits.

Our approach to optimization has two parts.  
First, we consider dephasing effects due to the detuning dependence of the qubit excitation frequency, which we call \emph{charge dispersion}, borrowing terminology from the superconducting qubit literature.
Charge dispersion describes the sensitivity of the qubit frequency to charge noise fluctuations in the detuning and associated dephasing. 
In a spirit similar to the transmon superconducting qubit, \cite{schreierPRB08,kochPRA07} we reduce the charge dispersion of the quantum dot hybrid qubit by tuning the static tunnel couplings. 
Recent experiments with parameters approaching this optimal parameter regime has improved $X$-rotation coherence times from $33$ ns to $177$ ns. \cite{brandur15}

However, because detuning is both the dominant noise source\cite{dialPRL13,peterssonPRL10,buizertPRL08,huPRL06} and a drive parameter in the hybrid qubit, minimizing charge dispersion while maintaining fast gate speeds, necessary for high fidelity operation, is problematic.
We show that this problem can be circumvented by driving instead the tunnel coupling, which is  more efficient, and results in higher gate speeds and fidelities than driving the detuning.

Second, we numerically optimize the $X$-rotation fidelity as a function of the ac drive amplitude and detuning. 
In this optimization, we simulate qubit dynamics using an effective two-dimensional Hamiltonian, to which we apply the Bloch-Redfield master equation for driven qubits and analytic formulas for dephasing rates due to $1/f$ noise. \cite{jingPRA14,yanNATC13,ithierPRB05,smirnovPRB03,makhlinPRL04}
We consider both the case of detuning and coupling drive, and find that  $X$-gate  fidelities exceeding $99\%$ can be achieved in both cases, and fidelities of  $99.8\%$ can be achieved by driving the tunnel couplings.  This optimal fidelity agrees with the result of simulations using a three state Hamiltonian that includes the nearest leakage state, averaged over numerically generated $1/f$ detuning noise. 

This paper is organized as follows.  
In section \ref{sec:qubit}, we summarize the relevant features of the quantum dot hybrid qubit.  
In section \ref{sec:chargedisp}, we explain our approach to finding the optimal static tunnel couplings that minimize charge dispersion.
In \ref{sec:acgates} and \ref{sec:decohere}, we describe our modeling of ac resonant gates and decoherence, respectively.
In section \ref{sec:fidelity}, we present the results of our numerical simulations on ac gate fidelities as a function of detuning and drive amplitudes.
A general procedure for deriving the effective Hamiltonian for a driven qubit is given in appendix \ref{app:canon}, and a summary of pure dephasing and relaxation rates due to $1/f$ noise, as well as a numerical procedure for numerically generating such noise, is given appendix \ref{app:noise}.

\section{Quantum dot hybrid qubit\label{sec:qubit}}

\begin{figure}[t]
\includegraphics[width=0.6\linewidth]{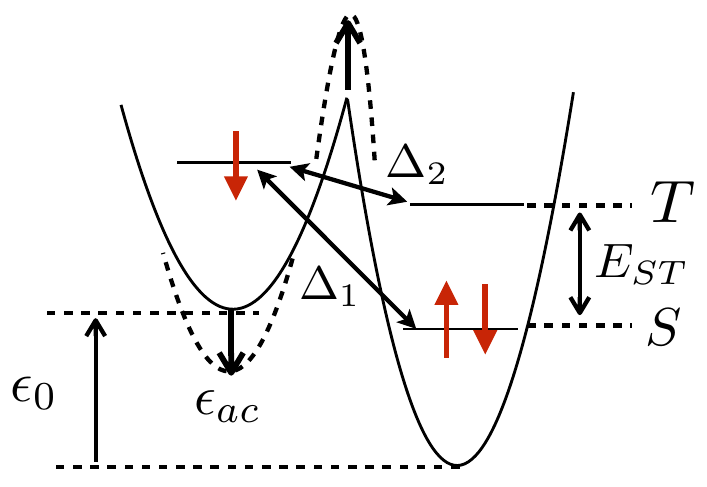}
\caption{(Color online) Illustration of the ac driven hybrid double quantum dot qubit.
The spin configuration of the state $\ket{{\rm\cdot}S}$ and the single dot energy levels corresponding to singlet and triplet state occupation on the right dot are shown. The electrostatic confinement potential is characterized by the detuning $\e_0$, the potential energy difference between the left and right wells, and the barrier height, which sets the tunnel coupling $(\D_1,\D_2)$.  
Detuning and tunnel coupling drive are implemented by modulation of the detuning ($\e_{ac}$) and the barrier height, respectively, shown in dashed lines.}
\label{DQD}
\end{figure}

The ``quantum dot hybrid qubit" \cite{kimNAT14,kohPRL12,shiPRL12} is formed by the  manifold of three-electron states in a double quantum dot with total spin quantum number ${S=1/2}$ and $z$-projection ${S_z=-1/2}$.
The three-electron double dot Hubbard Hamiltonian is given in Ref.~\onlinecite{shiPRL12}. 
A basis for the relevant states in the hybrid qubit regime can be chosen as  ${|{ \rm \cdot} S\>\equiv|{\rm \down}S\>}$, ${|S { \rm \cdot} \>\equiv|S{\rm \down}\>}$, and ${|{ \rm \cdot} T\>\equiv\sqrt{1/3}|{\rm \down} T_0\>+\sqrt{2/3}|{\rm \up}T_-\>}$, where $S(T)$ refers to singlet (triplet) states on one dot. 
In our notation, ${|{\rm \down}S\>=|{\rm \down}\>_L|S\>_R}$  denotes  spin down electron on the left dot and a singlet on the right dot, and  ${|{\rm \up}T_-\>=|{\rm \up}\>_L|T_-\>_R}$, etc. 
{The spin states of $|{ \rm \cdot} S\>$ and $|{ \rm \cdot} T\>$ are identical to those of the exchange-only logical qubit, \cite{diVincenzoNAT00} which consists of three electrons in a triple dot, instead of a double dot.}
This qubit is operated typically at an electron temperature of  $T=140$ mK.\cite{kimNATL15}

\begin{figure}[t]
\includegraphics[width=.7\linewidth]{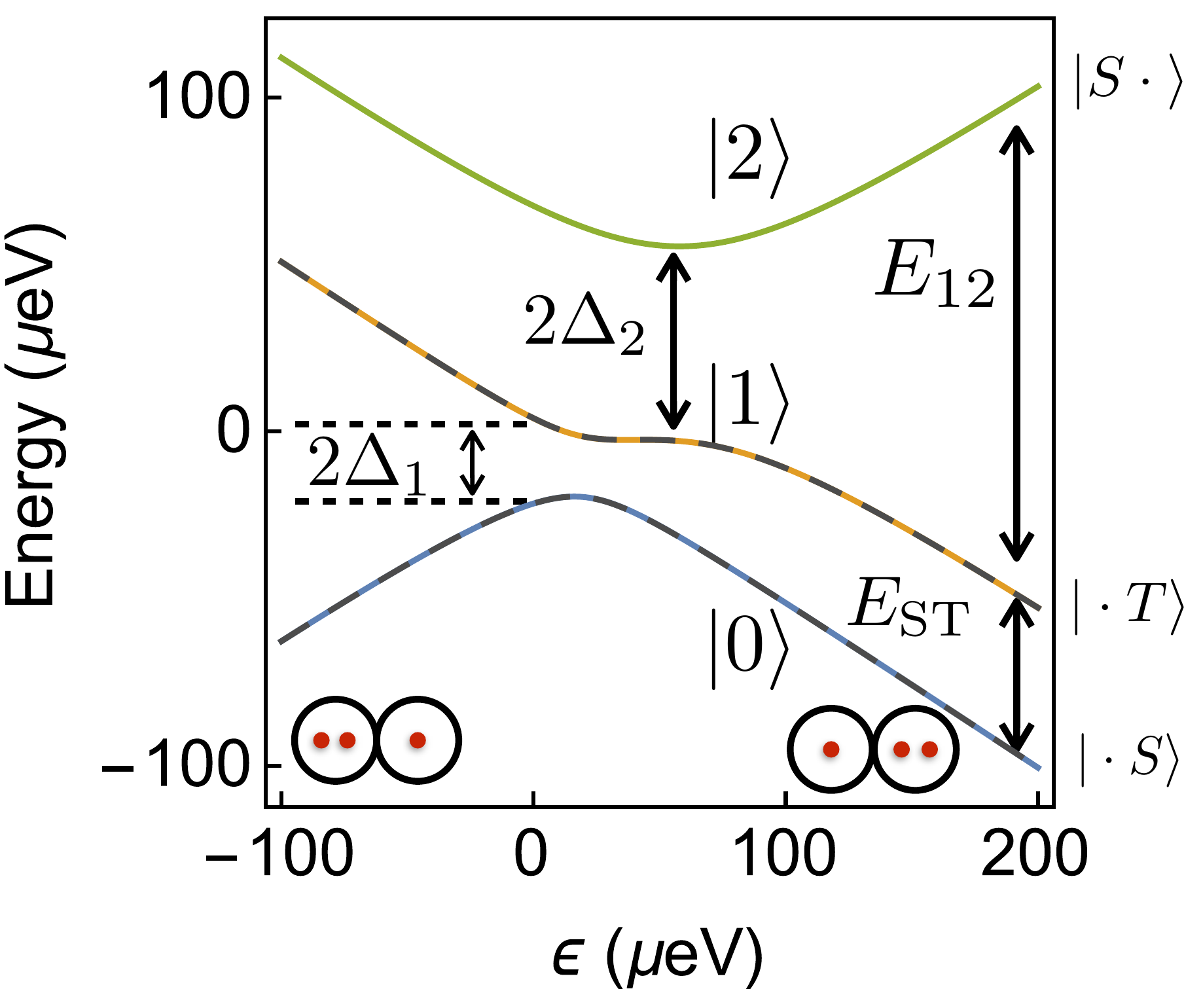}
\caption{(Color online) Qubit energy levels $E_0$ and $E_1$ and the leakage energy level $E_2$ as a function of detuning $\e$ computed with the three state Hamiltonian in Eq.~\eqref{eq:H3d} for parameters ${(E_{\rm ST},\D_1,\D_2)=(50,10.8,30.6)}~\mu$eV taken from Ref.~\onlinecite{kimNATL15}.
Black dashed lines: qubit energy levels computed from the static effective Hamiltonian Eq.~\eqref{eq:Heffdc}}
\label{fig:energy}
\end{figure}

Fig.~\ref{DQD} illustrates the basis states  $\{|{ \rm \cdot} S\>,|{ \rm \cdot} T\>,|S { \rm \cdot} \>\}$ in a gate-defined electrostatic potential.  The control parameters of the quantum dot hybrid qubit are the double dot detuning $\e$, defined as the energy difference between $(2,1)$ and $(1,2)$ charge states,  and the tunnel coupling $\D_{1}$  ($\D_2$) which cause charge transitions between $|S { \rm \cdot} \>$ and $|{ \rm \cdot} S\>$ ($|S { \rm \cdot} \>$  and $|{ \rm \cdot} T\>$).  
The Hamiltonian in the basis $\{|{ \rm \cdot} S\>,|{ \rm \cdot} T\>,|S { \rm \cdot} \>\}$ is given by
\ben
\mcal{H}(\e,\D_1,\D_2)=-{\e\over2}+
\begin{pmatrix}
0 & 0 & \D _1 \\
 0 & E_{ST}& -\D _2 \\
 \D_1 & -\D _2 & {\epsilon } \\
\end{pmatrix}\,,
\label{eq:H3d}
\een
where $E_{ST}$ is the energy splitting between the lowest lying singlet and triplet states on the right dot.
The energy levels of the Hamiltonian \eq{eq:H3d}, denoted by $\ket{0}$, $\ket{1}$ and $\ket{2}$, ordered from low to high energy,  are plotted  in Fig.~\ref{fig:energy} as  function of detuning, for the parameters ${(E_{\rm ST},\D_1,\D_2)=(50,10.8,30.6)}~\mu$eV taken from Ref.~\onlinecite{kimNATL15}.
The qubit logical states are $\ket{0}$ and $\ket{1}$,  while $\ket{2}$ is a leakage state.
We will represent the qubit on a Bloch sphere where $\ket{Z}=\ket{1}$ and $\ket{{\rm-}Z}=\ket{0}$ are located on the north and south pole, respectively.  
In this work, we consider operations of this qubit in the $(1,2)$ charge regime, $\e>\D_1$ and $\e-E_{ST}>\D_2$, where it is useful to think of the qubits states as mainly comprising of singlet-triplet states on the right dot, $\ket{0}\simeq|{ \rm \cdot} S\>$ and $\ket{1}\simeq|{ \rm \cdot} T\>$, with a small hybridization of the $|S{ \rm \cdot} \>$ state. 
 
% \mf{add minus sign in Eq. (2)}

The tunneling couplings $\D_1,\D_2$ generally have exponential dependence on detuning \cite{rastelliPRA12,shiPRL12}.  
In order to fit the experimentally observed resonant frequencies in Ref.~\onlinecite{kimNATL15}, we find it necessary to introduce such a dependence into the second tunnel coupling as 
\[{\D_{2}\to\til{\D}_2^0(\e)= \D^0_{2}e^{-|\e-E_{ST}|/\e_2}}\,,\]
 with ${\e_2=400~\mu}$eV, while $\D^0_1$  has a sufficiently long exponential decay length that it can be regarded a constant independent of $\e$.  This detuning dependence is included in Fig.~\ref{fig:energy} and in all results reported in this paper.  

\section{Optimal static tunnel couplings \label{sec:chargedisp}}

\begin{figure}[t]
\includegraphics[width=.8\linewidth]{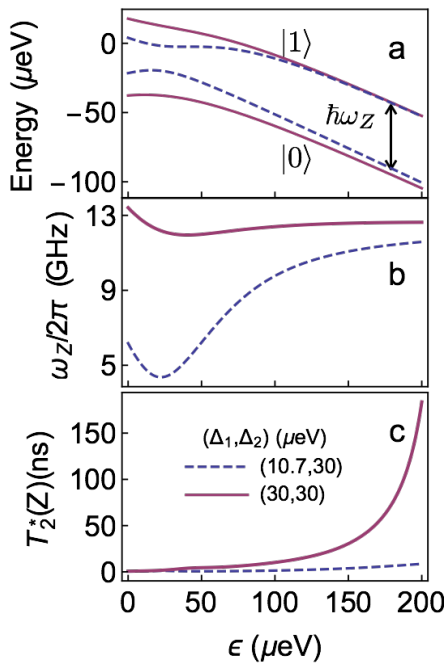}
\caption{(Color online) As function of detuning $\e$: (a) qubit energy levels $E_0$ and $E_1$, (b) qubit frequency $\w_Z/2\pi=E_{10}/h$, and (c) the quasistatic dephasing time scale $T_2^*$, for $(\D^0_1,\D^0_2)=(10,30)~\mu$eV (Dashed, blue) and $(\D^0_1,\D^0_2)=(30,30)~\mu$eV (Solid, Maroon).}
\label{fig:energy2}
\end{figure}

In this section, we tune the static tunnel coupling to minimize charge dispersion, defined as the detuning dependence of the qubit frequency ${\w_Z/2\pi=E_{10}/h}$, where ${E_{10}=E_1-E_0}$.  In the first instance, this will suppress the pure dephasing rate of free qubit precession  ($Z$ rotations) due to quasistatic detuning noise, given to leading order by ${\sqrt{2}/T_2^*(Z)=\s_\e^*|\p \w_Z/\p\e|}$, where\cite{makhlinCP04}
\ben
\s^*_\e= c_\e\sqrt{\frac{1}{\pi}\ln\pfrac{c_\e}{\hbar\w_l}}=5.7~\mu {\rm eV}
\label{sigstar}
\een 
 is an effective quasistatic noise variance, ${c_\e=2.38~\mu {\rm eV}}$ is the parameter in the $1/f$ detuning noise spectral density $S_\e(\w)=c_\e^2/\w$ [c.f.~\eq{eq:noisepower}], determined from experimental data on $T_1$ relaxation, as described in appendix \ref{sec:T1fit}, and  $\w_l\approx1$ Hz is the low frequency cutoff for the $1/f$ noise.
This $\s^*_\e$ value  is consistent with the static noise variance previously used to model the experiment of Ref.~\onlinecite{kimNAT14}.  
 A detail discussion of  pure dephasing rates due to quasistatic noise, including logarithmic corrections characteristic of the $1/f$  noise, is given in appendix \ref{sec:dephase1} and \eq{sigstar} is derived in appendix \ref{sec:lfnoise}.

At the same time, minimizing charge dispersion improves the coherence time of ac-driven $X$ rotation, which also suffers quasistatic dephasing from low frequency fluctuations of the Rabi frequency due to charge dispersion at quadratic order $\de\w_Z^2$, where $\de\w_Z=\de\e\p\w_Z/\p\e$, $\de\e$ denotes detuning noise, as discussed in section \ref{sec:lfdephase}.  
$Z$ rotations are also important because they can be combined with ac driven $X$ rotations to achieve universal single qubit control.

As evident from the energy level diagram, deep in the (1,2) charge regime ($\e\gg\D_1,\D_2$), the qubit frequency is insensitive to the detuning. 
Charge dispersion is minimal in this regime  because here logical states differs mainly in their \emph{spin} instead of charge character.
As a result, free precession is protected from dephasing due to charge noise fluctuations in the detuning.
A recent experiment in this regime demonstrated free induction decay at a frequency of $11.5$ GHz with a dephasing time of 10 ns, resulting in a $Z$-gate fidelity of 96$\%$.\cite{kimCM15}
We next show that this fidelity can be significantly improved by tuning the static tunnel couplings.

To find the optimal static tunnel couplings parameter regime, consider the following qualitative argument.
Charge dispersion comes from level repulsion at the anticrossings due to the tunnel couplings.  Specifically, as one moves  towards $\e=0$ from large detunings $\e\gg\D_1$, the energy level $E_1$ is repelled downwards at  ${\e=E_{ST}}$ by  $\D_2$ and then repelled upwards  at ${\e=0}$ by $\D_1$, see Fig.~\ref{fig:energy}.
We thus expect the net effect of level repulsion to be minimized when $\D^0_1\simeq \D^0_2$.  By performing an empirical search, we find minimal charge dispersion at the tunnel couplings ${\D^0_1=\D^0_2=30~\mu}$eV.
Fig.~\ref{fig:energy2} a--c compares the qubit energy levels, excitation frequency $\w_Z/2\pi$, and the quasistatic dephasing time scale $T_2^*(Z)$ for the optimal tunnel couplings with the ones reported in Ref.~\onlinecite{kimCM15}.   
The curvature in the energy levels near the anticrossing and associated charge dispersion is greatly reduced, and, as a result, $T_2^*(Z)$ is increased by an order of magnitude  from $10$ ns to $180$ ns.

%In addition to dephasing due to quasistatic noise
In principle, the high frequency  components of the $1/f$ noise spectrum can also cause dephasing, due to quadratic charge dispersion ${\de\e^2\p^2\w_Z/\p\e^2}$.\cite{makhlinPRL04}  The resulting decay envelop is exponential with a decay time $\tau_2$ inversely proportional to the curvature of the charge dispersion  $\tau_2(Z)\propto (\p^2\w_Z/\p\e^2)^{-1}$.  However, this decay is strongly suppressed for the optimal tunnel couplings, as $\tau_2>1~\mu$s in the far detuned regime, as shown in appendix \ref{sec:hfnoise}.  Since the the $Z$-gate times are about $t_g(Z)\sim0.1$ ns, this decay should lead to a $Z$-gate infidelities less than $t_g(Z)/\tau_2(Z)\simeq0.01\%$, which is very small.

Henceforth, we set static tunnel couplings at their optimal values, and optimize ac $X$ gates as a function of remaining parameters.

\section{AC resonant gates\label{sec:acgates}}
Ac resonant techniques for producing transitions between the logical qubit states ($X$ rotations) are similar to electron spin resonance (ESR), where the logical qubit plays the role of the electron spin  and oscillations at the qubit frequency $\w_Z/2\pi$ in the electrostatic control parameters plays the role of the resonant driving field.  Even in the absence of decoherence, several complications arise in ``logical qubit resonance" (LQR) \cite{kohPNAS13} which do not occur in ESR.
In implementing LQR, one has indirect control over the direction and magnitude of the driving field, which is determined by the qubit response to perturbations in the control parameters.     
Furthermore, to achieve fast gate speeds, one may have to enter the strong driving regime where the qubit's nonlinear response can spoil gate fidelity.
While this problem could be addressed by properly shaping the driving pulse, as demonstrated with diamond NV centers,\cite{MotzoiPRL09,gambettaPRA11,fuchsSCI09} 
we do not consider these tecniques in this paper.

LQR for the hybrid qubit is performed by adiabatically initializing to a detuning $\e_0$, and then applying ac oscillations of the detuning (voltage bias between the quantum dots)
\begin{align}
\e(t)&=\e_0+\e_{ac}(t)\,,\nn
\D_1(t)&=\D_1^0\nn
\D_2(t)&={\D}^0_2e^{-(\e(t)-E_{ST})/\e_2}\,,
\end{align}
or tunnel couplings (DQD barrier height), 
% which are described by parametrizations,
\begin{align}
\D_1(t)&=\D_1^0+\D_{ac}(t)\nn
\D_2(t)&=\til{\D}_2(\e_0)+\D_{ac}(t)\,,
\end{align}
where we will consider the specific drive signals,
\[ \begin{pmatrix}\e_{ac}(t)\\ \D_{ac}(t)\end{pmatrix}=\cos\w_Z t\begin{pmatrix}A_\e\\A_\D\end{pmatrix}\,,\quad \w_Z=E_{01}/\hbar\,.\]
where $A_\e$ and $A_\D$ are the drive amplitudes of detuning and coupling drive, respectively.
These two types of driving are illustrated in Fig.~\ref{DQD}.
Note that, due to the energy dependent tunnelling, detuning drive will also result in an ac signal in the second tunnel coupling, as well.
The total Hamiltonian now comprises of a dc and ac component,  ${\mcal{H}(t)=\mcal{H}_{dc}+\mcal{H}_{ac}(t)}$, where $\mcal{H}_{dc}=\mcal{H}(\e_0,\D_{1}^0,\D_2^0)$ is the dc component,  and 
\[\mcal{H}_{ac}(t)=\mcal{\mcal{H}}[\e(t),\D_{1,2}(t)]-\mcal{H}_{dc}\]
 is the ac component.  

Tunnel coupling driving should enable faster, more efficient (fast  speed per drive amplitude), and higher fidelity gates than detuning drive, because of stronger transition matrix elements between qubit logical states at the large detunings where the qubit is protected from dephasing.  There,  charge hybridization between $(1,2)$ and $(2,1)$ are minimal,  so detuning drive, which do not couple definite charge states, cannot be effective, while transitions between $\ket{0}\simeq\ket{\cdot S}$ and $\ket{1}\simeq\ket{\cdot T}$, mediated by tunneling (see Fig.~\ref{DQD}), have matrix elements of the order $\D_1\D_2/\e$.
This drawback of detuning drive is quite generic: since the gate speed  scales as the double dot susceptibility to detuning fluctuations, increasing drive efficiency also increases sensitivity to detuning noise. 
This is particularly important for finite frequency noise, because the rotating frame pure dephasing rate $1/T_{\phi}'$ for ac $X$ rotation scales quadratically with the detuning drive Rabi frequency, as shown in section \ref{sec:acBloch},  \eq{eq:Tphi}.  
% so they can only be fast when operated in a charge qubit regime, 
%This means one has to operate near charge transition to achieve fast gate speeds, where charge relaxation is strong.
This problem is evident in the experiment of Ref.~\onlinecite{kimNATL15}, where  high speed ($\sim1$ GHz) ac gates were achieved at detunings near the $(1,2)$ to $(2,1)$ charge transition, but the decay times were only 1-2 ns, even at the charge noise sweet spot,  limiting gate fidelities to $\sim86\%$. 

To model ac gates, it is more convenient to express the Hamiltonian in the basis of energy eigenstates, ${\til{\mcal{H}}(\e_0)=\mcal{U}^\dag(\e_0)\mcal{H}\mcal{U}(\e_0)}$, where $\mcal{U}$ is the unitary transformation between eigenstates $\{\ket{n}\}$ of the static Hamiltonian at $\e_0$ and ${\{|{ \rm \cdot} S\>,|{ \rm \cdot} T\>,|S { \rm \cdot} \>\}}$.
Using this Hamiltonian, we can estimate the leakage probabilities for short times from qubit states $\ket{m}$ to the leakage state $\ket{2}$  by ${p_{m\to2}\simeq(\til{\mcal{H}}_{m2}/E_{2m})^2}$, where ${m={0,1}}$ and ${E_{2m}=E_2-E_m}$ are the energy gaps to the leakage state.
The leakage is very small at large detunings, $<0.1\%$ for ${\e>200~\mu}$eV, due to the large energy gap to the leakage state, ${E_{2m}>150~\mu}$eV.  We have also verified this leakage estimate by numerical solution of the density matrix  equation using the three-state Hamiltonian Eq.~\eqref{eq:H3d}. 
Since the leakage is negligible, the qubit dynamics is governed by an effective two dimensional Hamiltonian, presented in section \ref{sec:Heff}. 

\subsection{Effective Hamiltonian for the driven qubit\label{sec:Heff}}
In this section, we analyze the qubit dynamics and noise using the effective Hamiltonian for the driven qubit, which is useful for both numerical and analytical calculations, for providing intuition, and for applying techniques of electron spin resonance to the driven qubit.   Applying standard techniques\cite{foldyPR50,winklerS0Book03} described in appendix \ref{app:canon}, we find an effective Hamiltonian given by  a perturbative expansion in the inverse energy gaps between the qubit and the leakage state, $E^{-1}_{21}$ and $E^{-1}_{20}$, and can generally be written as 
\ben
h=\mbb(\e_0,\e_{ac}(t)+\de\e(t),\D_{1,2}(t))\cdot\bm{\s}/2\,,
\label{eq:heff}
\een
where $\de\e(t)$ is the detuning noise,  $\bm{\s}$ are the Pauli matrices,  $\mbb$ is the effective magnetic field that is a nonlinear function of the qubit control parameters $(\e,\D_1,\D_2)$.
We retain the leading order term in the effective Hamiltonian, given in \eq{appeq:h}. 
{The static effective Hamiltonian $h_{dc}$, given in \eq{eq:Heffdc}, was derived in Ref.~\onlinecite{friesenCM15}, and the qubit energy levels computed with it, shown in \fig{fig:energy}, agrees very well with the ones calculated with the static three-state Hamiltonian $\mcal{H}_{dc}$ [\eq{eq:H3d}]}.
{However, $h$ generally has a different functional dependence on $\e_0$ and $\e_{ac}$, as indicated in \eq{eq:heff}, due to $\e_0$-dependent transformations used to approximately block diagonalizes the static Hamiltonian ($\mcal{H}_{dc}$) between qubit and leakage states. 
These transformations define the basis in which $h_{dc}$ is derived, which is related to the basis $\{|{ \rm \cdot} S\>,|{ \rm \cdot} T\>,|S { \rm \cdot} \>\}$ of Eq.~\eqref{eq:H3d} by Eq.~\eqref{eq:basis}. 
In particular, we emphasize that one cannot derive \eq{eq:heff} by simply taking $\e_0\to\e_0+\e_{ac}$ in $h_{dc}$. }
 
In the far detuned limit ($\e\to\infty$), the matrix elements of \eq{eq:heff} are given by 
\begin{align}
h_{00}&=-\frac{\epsilon }{2}-\frac{\Delta _1^2}{\epsilon}\,,\quad h_{11}=E_{ST}-\frac{\epsilon }{2}+\frac{\Delta _2^2}{E_{ST}-\epsilon}\,,\nn
h_{01}&=\frac{\Delta _1 \Delta _2}{2} \left(\frac{1}{\epsilon }-\frac{1}{E_{ST}-\epsilon }\right)\,.
\label{eq:h1}
\end{align}
The quadratic terms describe virtual transitions to the leakage state, which would otherwise appear as second order terms in the time evolution operator for the 3D Hamiltonian Eq.~\eqref{eq:H3d}.   \cite{sakuraiQM,kittelQTS87}

\begin{figure}[t]
{\centering}
{\includegraphics[width=\linewidth]{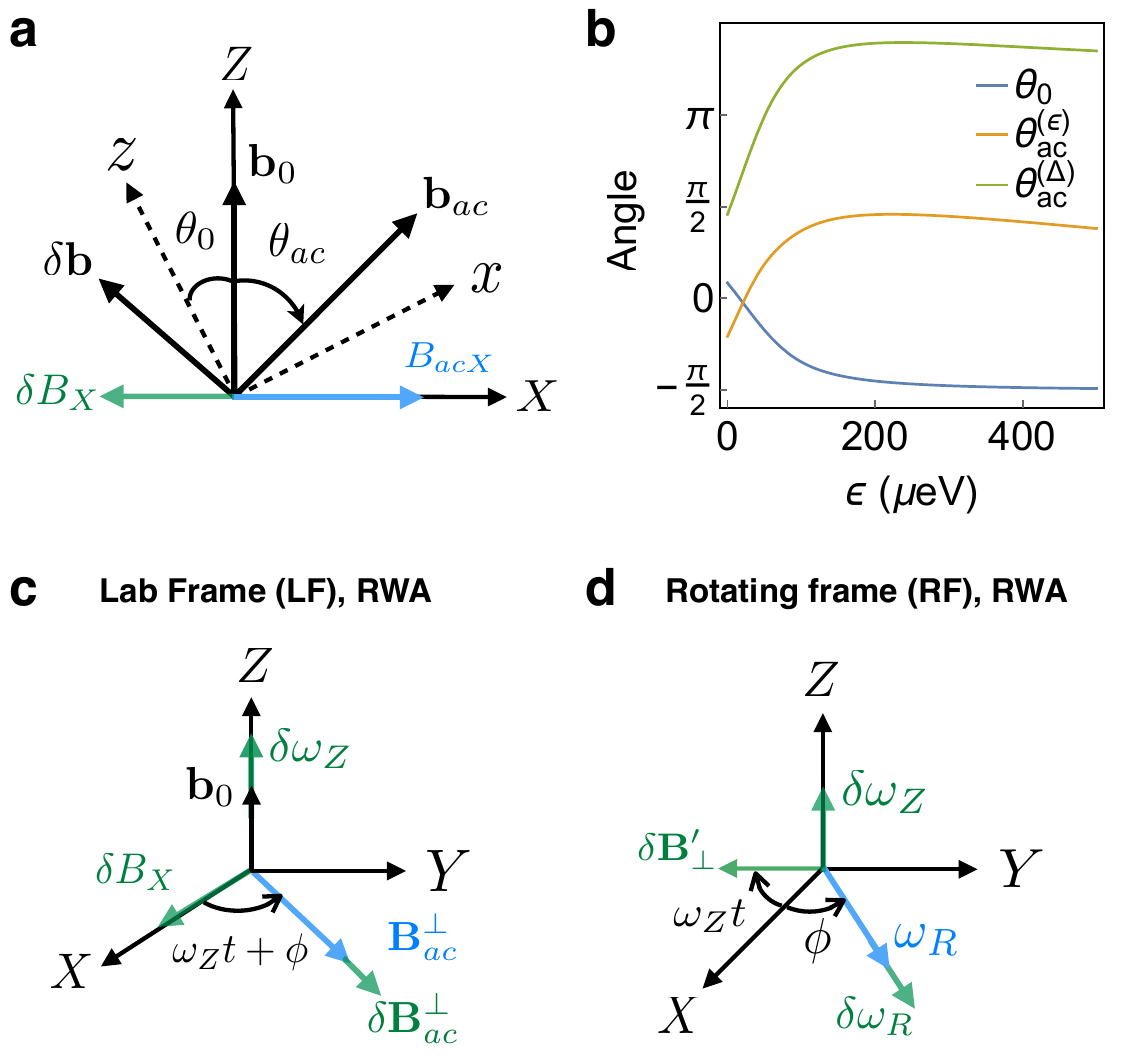}}
\caption{(Color online)
(a) At a fixed detuning $\e_0$, ${X-Z}$ axes of the Bloch sphere, representing the qubit logical basis, and the $x-z$ axes, representing the basis in which the effective Hamiltonian \eq{eq:h1} is derived.   The $Z$ axes lies along the static field $\mbb_0$, 
%which defines the local energy eigenstates, 
at an angle $\theta_0$ rotated from the $z$ axis.
The ac drive field $\mbb_{ac}$ lies at an angle $\theta_{ac}$ from the $Z$ axis, and has the transverse component $B_{acX}$ (blue arrow).
 $\de\mbb$ is the noise field with transverse component $\de B_X$, shown in green.
(b) Detuning dependence of the static field angle $\theta_0$ and the ac drive field angles for detuning ($\theta_{ac}^{(\e)}$)  and tunnel coupling ($\theta_{ac}^{(\D)}$) drive. 
(c and d) Effective fields in the rotating wave approximation when a phase $\phi$ is included in the drive amplitude, which goes as $\cos(\w_Zt+\phi)$.  
(c) Lab frame: the transverse resonant  component of the ac drive field $\mbB_{ac}^\perp$ (blue arrow) and the noise in its drive amplitude $\de\mbB^\perp_{ac}$ (green arrow) [\eq{eq:bac1}], is rotated at an angle $\w_Zt+\phi$ from the $X$ axis at time $t$.
The $X$ and $Z$ components of the noise field, $\de B_X$ and $\hbar\de\w_Z$ [\eq{Hac}], respectively, are shown in green.
(d) Rotating frame: 
%The static field $\mbb_0$ is absent. 
The transverse ac drive field is fixed at an angle $\phi$ and has magnitude $\hbar\w_R$ (blue arrow) with a noise component $\hbar\de\w_R$ (green arrow). 
The tranverse noise field $\de\mbB_\perp'$ (green arrow) [\eq{dbperp}] counter-rotates at the qubit frequency $\w_Z$.}
\label{fig:EffFields}
\end{figure}

The effective magnetic field Eq.~\eqref{eq:heff} has three components
\ben
\mbb(t)=\mbb_0+\mbb_{ac}(t)+\de\mbb(t)\,,
\label{eq:b}
\een
as illustrated in Fig.~\ref{fig:EffFields}a, where $\mbb_0=\mbb(\e_0)$ is the static, dc field corresponding to $\mcal{H}_{dc}$, which will define the ``longitudinal" direction in the lab frame,  $\mbb_{ac}$ is the ac field corresponding to $\mcal{H}_{ac}$, and $\de\mbb$ is the effective field due to noise, defined by $\mbb(\e_0+\de\e)-\mbb(\e_0)$.
We can generally parametrize the static field as
\[{\mbb}_0=\sin\theta_0\hatbf{x}+\cos\theta_0\hatbf{z}\,,\]
where the detuning dependence of the angle $\theta_0=\theta(\e_0)$, along with the relative angles between the drive field $\mbB_{ac}$ and $\mbb_0$, for both types of driving, denoted by $\theta^{(\e)}_{ac}$ and $\theta^{(\D)}_{ac}$, are plotted in Fig.~\ref{fig:EffFields}b.
Unlike conventional spin resonance, $\mbb_{ac}$ has both longitudinal and transverse component, but only the transverse field $\mbb_\perp(t)=\hatbf{b}_0\times\mbb(t)\times\hatbf{\mbb}_0$ drives qubit transitions.    

We next present a general analysis based an expansion of the effective Hamiltonian up to second order in drive and noise fields  that  captures all the relevant physics in the parameter regime of interest.
The results reported below in section \ref{sec:fidelity}, however, follow from numerical simulations with the effective Hamiltonian \eq{appeq:h} that include higher order nonlinear effects.

The noise field due to fluctuations in the detuning $\e$ and the singlet-triplet energy splitting $E_{ST}$  is given by,
\ben
\de\mbb\simeq\de\e\frac{\p\mbb}{\p(\de\e)}+\de E_{ST}\frac{\p\mbb}{\p{E_{ST}}}\,.
\label{db}
\een
Note that this linear expansion will lead to second order noise correlations and associated relaxation terms in the density matrix master equation, see section \ref{sec:decohere}, which is derived by solving for the system-bath interaction to second order in perturbation theory.\cite{gardinerBook04}  
The driving field up to quadratic order is given by
\begin{align}
\mbb_{ac}&=\mbb_{ac}^{(1)}+\mbb_{ac}^{(2)}\nn
\mbb_{ac}^{(1)}&=u(t)\frac{\p\mbb}{\p u}+\de\mbb_{ac}\,,\quad
\de\mbb_{ac}=u(t)\de\e\frac{\p^2\mbb}{\p u\p(\de\e)}\label{eq:bac1}\\
\mbb_{ac}^{(2)}&=\frac{u^2(t)}{2}\frac{\p^2\mbb}{\p u^2}
\label{bac}
\end{align} 
where~$u=\e_{ac}~{\rm or}~\D_{ac}$, and $\mbb_{ac}^{(1)}$ and $\mbb_{ac}^{(2)}$ are the term linear and quadratic in $u$.  In the second equation, the term proportional to $\de\e$ represent fluctuations in the drive amplitudes due to detuning noise.

In the Bloch sphere representation of \eq{eq:heff}, the qubit logical states, defined by the eigenstates of  $\mbf{b}_0\cdot\bm{\s}$, lies along the axis defined by $\mbb_0$. 
It will be convenient to use instead a basis in which the logical states always lie along the $Z$-axis.
To this end, we apply the unitary transformation 
\[U_0=\begin{pmatrix}\cos(\theta_0/2)&-\sin(\theta_0/2)\\ \sin(\theta_0/2)&\cos(\theta_0/2)\end{pmatrix}\,,\]
which diagonalizes the static Hamiltonian, ${U^\dag_0(\hatbf{b}_0\cdot\bm{\s})U_0=\s_Z}$.  
We write the effective Hamiltonian in the local basis as 
\ben
H=U^\dag(\theta)hU(\theta)=\onehalf(B_Z(t)\s_Z+B_X(t)\s_X)\,,
\label{appeq:h}
\een
where ${B_i(t)=\mbb(t)\cdot\mbf{e}_i}$, $\{\mbf{e}_i\}$ are detuning-dependent qubit pseudospin basis vectors given by $\mbf{e}_Z(\e_0)=\hatbf{b}_0$, $\mbf{e}_X(\e_0)=\cos\theta_0\hatbf{x}-\sin\theta_0\hatbf{z}$, and $\mbf{e}_Y=\hatbf{y}$, and indices in capital letters denote  local frame axes.
These local axes are illustrated in Fig.~\ref{fig:EffFields}a.

Up to the expansions Eq.~\eqref{bac}, the effective field components in \eqref{appeq:h} are given by
\begin{align}
B_Z(t)&=-\hbar(\w_Z+\de\w_Z)+B^{(1)}_{acZ}+B^{(2)}_{acZ}\nn
B_X(t)&=2\hbar(\w_R+\de\w_R)\cos\w_Zt+\de B_X+B^{(2)}_{acX}\,.
\label{Hac}
\end{align}
where the Rabi angular frequency $\w_R$ and its fluctuation $\de\w_R$ are given by 
\ben
\w_R=\frac{A_u}{2\hbar}\frac{\p B_X}{\p u}\,,\quad \de\w_R=\de\e\frac{\p\w_R}{\p\de\e}\,.
\label{eq:rabi}
\een
The noise term $\de\w_R$ is a nonlinear effect that comes from second order processes in which $\e_{ac}$ drives a transition from the qubit subspace into the leakage state, and then noise $\de\e$ drives transition back into the qubit subspace, or vice versa, as illustrated in \fig{fig:drabi}a.

\begin{figure}[t]
\begin{center}
\includegraphics[width=.5\linewidth]{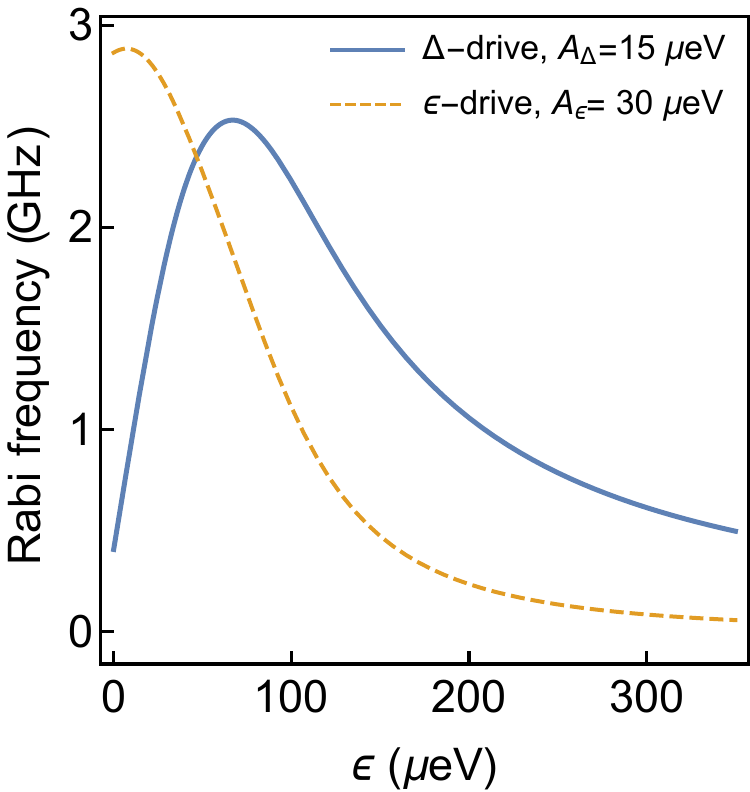}
\caption{(Color online) Rabi frequency as a function of detuning $\e$, computed from Eq.~\eqref{eq:rabi}, for detuning drive amplitude $A_\e=30~\mu$eV and coupling drive amplitude $A_\D=15~\mu$eV.}
\label{fig:rabi}
\end{center}
\end{figure}

The Rabi frequency ${|\w_R|/2\pi}$ is plotted for detuning drive with ${A_\e=30~\mu}$eV and coupling drive with ${A_\D=15~\mu}$eV in Fig.~\ref{fig:rabi}.  As expected from our argument in section \ref{sec:acgates}, tunnel coupling driving is more efficient and stronger at large detunings. 
On the other hand, fast detuning drive gate speeds can be attained at lower detunings where charge dispersion is larger. 
%%The relationship between gate speed and energy level curvature is given in \eq{eq:RabiDetune} of section \ref{sec:Heff}.
We can quantify this relationship by considering the scaling of the charge dispersion curvature with the detuning drive efficiency $\p B_X/\p\e_{ac}=2\hbar\w_R/A_\e$.
From \eq{eq:rabi}, \eq{eq:dE}, and \eq{appeq:a12Z}, we find
\ben
\frac{\p^2E_{10}}{\p\e^2}\propto\frac{2\hbar(\w_R/A_\e)^2}{\w_Z}\label{eq:RabiDetune}\,,
\een
where we used $\p\mbb/\p(\de\e)=\p\mbb/\p\e_{ac}$,\footnote{Note that  because the basis vectors $\mbf{e}_i$ do not depend on the driving parameters $u$ or noise $\de\e$,  we have $\p B_i/\p u=\mbf{e}_i\cdot{\p \mbb}/{\p u}$, and $\p B_i/\p \de\e=\mbf{e}_i\cdot{\p \mbb}/{\p\de\e}$.}  see \eq{eq:heff}, and neglected terms that are independent of $\w_R$ which comes from $\p^2 B_Z/\p\e_0^2$  and the difference between $\p\mbb/\p\e_0$ and $\p\mbb/\p\e_{ac}$, see \eq{eq:deps} in appendix \ref{app:canon}.
The charge dispersion thus scales quadratically with $\w_R$. 

\subsection{Rotating wave approximation\label{sec:RWA}}
While we will simulate the unitary qubit dynamics governed by the full effective Hamiltonian in section \ref{sec:fidelity}, it will be instructive to analyze the qubit dynamics in the rotating wave approximation (RWA). The leading corrections to the RWA are of order $\w_R/\w_Z$, see, e.g.~, Ref.~\onlinecite{smirnovPRB03}.
The rotating frame (RF) Hamiltonian, defined by 
\ben
H_{RF}=-i\hbar U_Z^\dag\p_tU_Z +U_Z^\dag HU_Z
\label{eq:HRF}
\een where $U_Z$=$e^{i\w_Z t\s_Z/2}$,  is given in the RWA by
%discuss application of Bloch relaxation equations from schllicter's book 
\begin{align}
H_{RF}^{(RWA)}&={\hbar\over2}[(\w_R+\de\w_R)\s_X+\de \w_Z\s_Z]+\de\mbB_\perp'\cdot{\bm{\s}\over2}\nn
&+B_{acZ}^{(1)}\s_Z+\mbB_{ac}^{(2)}\cdot{\bm{\s}\over2}
\label{eq:HrfRWA}
\end{align}
where the transverse noise field has additional time dependence (indicated by $'$) corresponding to $Z$ rotations at the driving frequency $\w_Z$
\ben
\de\mbB_\perp'(t)=\de B_X(t)(\cos\w_Zt\mbf{e}_X-\sin\w_Zt\mbf{e}_Y)\,,
\label{dbperp}
\een
while $\de\w_Z$ and  $\de\w_R$ are unaffected by the transformation. 
Note that the addition of a  phase $\phi$ in the ac drive signal  ($\cos \w_Zt\to\cos(\w_Zt+\phi)$) enables rotations in the RF about an axis in the $XY$ plane at an angle $\phi$ rotated from the $X$ axis, as illustrated in Fig.~\ref{fig:EffFields}d.
Since only two axis control of the qubit is required for universal gate operations, this qubit could be operated entirely with ac gates.
Fig.~\ref{fig:EffFields}d illustrates the relevant effective fields in the rotating frame.

In addition to the standard spin resonance Hamiltonian in the rotating frame, Eq.~\eqref{eq:HrfRWA} has a  longitudinal drive $B_{acZ}$ and a nonlinear drive term $\mbB_{ac}^{(2)}$.
The longitudinal driving term does not affect our results because we operate the qubit in the regime $\w_R\simeq\w_Z$.\cite{glennPRB13}
Details about the magnitude and effects of this term are given in appendix \ref{LdriveApp}.  
The second harmonic of the driving field is given by
\[\mbB_{ac}^{(2)}=A_{u}^2\frac{1+\cos2\w_Zt}{4}\frac{\p^2\mbB}{\p u^2}\,.\]
The ac term with frequency $2\w_Z$ is off-resonant, so it can be neglected consistently in the RWA.
On the other hand, the dc term drives rotations at a frequency set by ${A_u^2(\p^2\mbb/\p u^2)/4}$, which can cause beating with oscillations driven by the linear transverse ac driving field $B_{acX}$.
This beating pattern can be seen in oscillations of the infidelities as a function of $\e$  for large drive amplitudes, as seen in \fig{infidel}d of section \ref{sec:acBloch}.

\section{decoherence\label{sec:decohere}}
In the following, we apply the decoherence model for the driven qubit in the presence of $1/f$ noise, previously developed for superconducting qubits \cite{yanNATC13,ithierPRB05,makhlinPRL04,smirnovPRB03}, to the quantum dot hybrid qubit.
In section \ref{sec:noisepower}, we discuss the relevant noise sources and their power spectrum. 
In section \ref{sec:mitlfnoise}, we show how ac driving avoids the low frequency noise in $\de B_X$.
In \ref{sec:acBloch}, we apply the Bloch-Redfield equations in the rotating frame to the driven qubit, and compute the associated relaxation rates.
In \ref{sec:lfdephase}, we estimate the effect of the low frequency noise in the Rabi frequency $\de\w_R$.
The results derived from our decoherence model are checked with simulations using the three state Hamiltonian with numerically generated $1/f$ detuning noise in section \ref{sec:fidelity}. 

\subsection{Noise sources, power spectrum, and relaxation tensors\label{sec:noisepower}}
As we mentioned in the introduction, charge noise with $1/f$ type noise spectrum is the dominant cause of decoherence for double quantum dots in Si.
Electrostatic coupling to this charge noise causes fluctuations in the detuning  $\de\e$ and the singlet triplet splitting $\de E_{ST}$, which are completely characterized by the classical noise autocorrelation 
\begin{align}
S_{\epsilon} (t-t')&=\langle\delta\epsilon(t)\delta\epsilon(t')\rangle\nn
S_{ST} (t-t')&=\langle\delta E_{ST} (t)\delta E_{ST} (t')\rangle\,.
\label{S}
\end{align} 
Their noise power spectrum $S(\w)=\int\,dte^{-i\w t}S(t)$ are given by\footnote{The noise strengths coefficients $c_\e$  and $c_{ST}$ are proportional to the temperature \cite{culcerAPL09}.}
\begin{align}
S_\e(\w)&=\frac{c^2_\e}{|\w|}\,,\quad 
S_{ST}(\w)=\frac{c^2_{ST}}{|\w|}\,\quad
\w_l<|\w|<\w_h\,,
\label{eq:noisepower}
\end{align}
where we impose a sharp high and low frequency cutoff, $\w_h$ and $\w_l$, respectively.  The low frequency cutoff $\w_l$ is set by the total measurement time ${\w_l=2\pi/\tau_m}$ \cite{ithierPRB05,cywinskiPRB08}, which is ${\tau_m=200}$ ms in the present experiment, corresponding to a low frequency cutoff of $\tau_m^{-1}=5$ Hz.\footnote{We assume here that the physical microscopic cutoff is not higher than $\w_l$.}
% while the correlation time is approximately given by $\tau_c\sim \tau_m$.
%Because $T_1$ relaxation between for the hybrid qubit state which requires processes that cause transition between different spin states  is negligibly small \cite{pradaPRB08}.

\begin{figure}[t]
\begin{center}
\includegraphics[width=\linewidth]{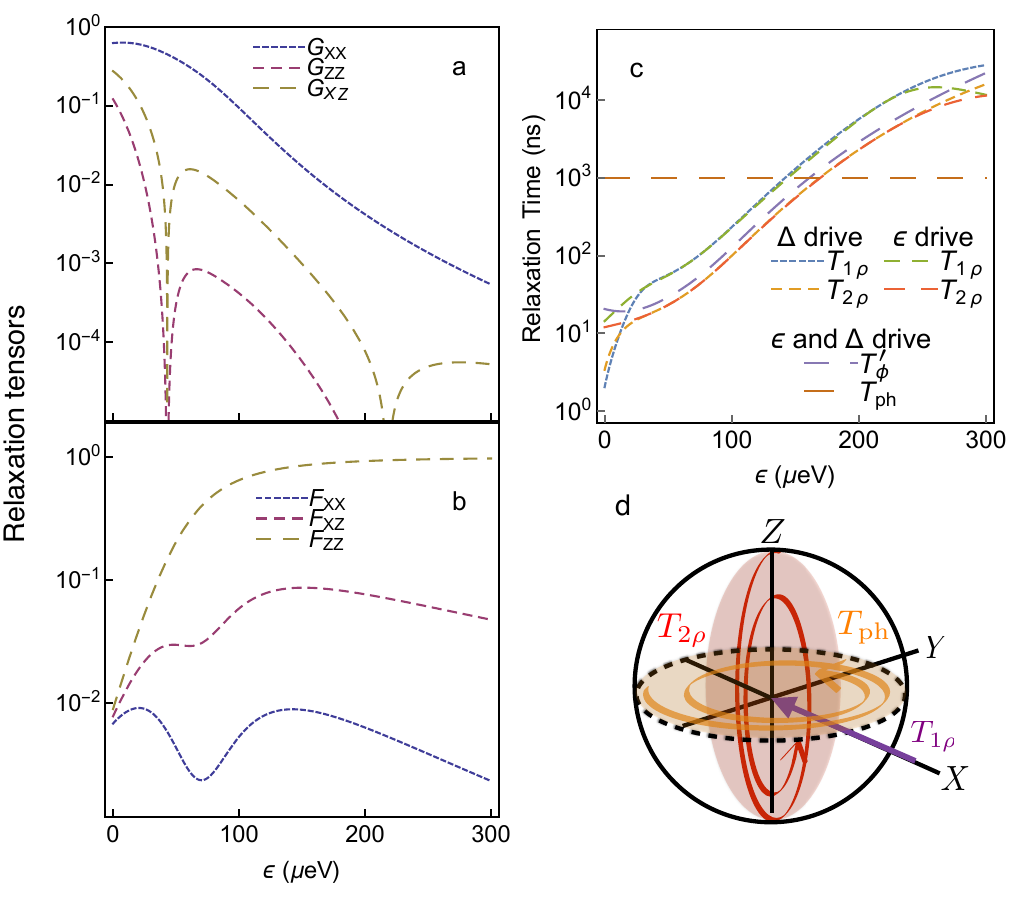}
\caption{(Color online)
For ${(\D^0_1,\D^0_2)=(30,30)~\mu}$eV and as a function of detuning $\e$: (a and b) The relaxation tensors \eq{GFtensors}  (a)  $G_{ij}$  for detuning noise and (b) $F_{ij}$ for noise in the singlet-triplet splitting $E_{ST}$. (c) For the drive amplitudes ${(A_\e,A_g)=(30,10)~\mu}$eV,   relaxation times $(T_\phi',T_{1\rho},T_{2\rho})$ Eq.~\eqref{T2rf} in the Bloch equations for both types of ac drive, and the phonon relaxation time $T_{\rm ph}$. 
(d) Illustration of the longitudinal and transverse relaxation process in the rotating frame, relative to the drive field ($\w_R$), governed by $T_{1\rho}$, and $T_{2\rho}$, respectively, and the phonon induced relaxation ($T_{\rm ph}$).
%that causes decay of the singlet-triplet coherence.
   }
\label{fig:relax}
\end{center}
\end{figure}

Dephasing and relaxation of qubit dynamics,  described by the master equation in section \ref{sec:acBloch}, is not determined directly by the noise power in \eq{eq:noisepower}, but the power spectrum of the noise field that appears in the effective Hamiltonian \eq{eq:heff}, given by \footnote{We neglect here any correlations between $ \de E_{ST}$ and $\de\e$.}
\begin{align}
S_{ij}(\w)&=\<\de B_i(\w)\de B_j(-\w)\>=G_{ij}S_\e(\w)+F_{ij}S_{ST}(\w)
\label{S}
\end{align}
where we define the relaxation tensors
\ben
G_{ij}(\e)=\frac{\p B_i}{\p(\de\e)}\frac{\p B_j}{\p(\de\e)}\,,\quad F_{ij}(\e)=\frac{\p B_i}{\p E_{ST}}\frac{\p B_j}{\p E_{ST}}\,.
\label{GFtensors}
\een
These tensors, which are anisotropic and detuning dependent, describe the qubit susceptibility to fluctuations in $\e$ and $E_{ST}$.
They are plotted as a function of detuning in Fig.~\ref{fig:relax} a and b.  
The decrease in $G_{ij}(\e)$ as a function of detuning is consistent with our discussion in secion II and III: when the logical states have the same charge character, $\de\e$ cannot cause changes in the qubit frequency or  transitions between logical states.
% which depend on qubit parameters under experimental control, determine 
On the other hand, as the qubit logical states acquire different spin character as $\e$ increases, they
become more sensitive to $\de E_{ST}$ noise, so that  $F_{ij}$ increases.

%Our qubit optimization corresponds to choosing the magnitude and direction of the noise field, as dicted by these tensors, detunings where

We will next make some approximations to $S_{ij}$ appropriate for the optimal parameters considered in this work.
First, detuning noise is by far the dominant noise parameter, with ${c_\e=2.38~\mu {\rm eV}}$, while an estimate in Ref.~\onlinecite{gamblePRB12} gives ${c_{ST}\sim10^{-3}~\mu}$eV.\ {The large difference in noise strength occurs because the detuning couples to charge noise  via the double dot electric dipole moment, but while the single-triplet splitting $E_{ST}$ couples via a single dot quadrapole moment.\cite{gamblePRB12}}
Since even the largest relaxation rate associated $\de E_{ST}$, set by the noise power at the $\w_Z$ or $\w_R$ as discussed in section \ref{sec:acBloch}, is given by $F_{ZZ} c_{ST}^2/\hbar^2\w_R\simlt100$ Hz,  negligibly small compared to  other relaxation rates considered in this work, we can neglect $\de E_{ST}$ altogether in the decoherence of ac gates.\footnote{This is true for ac gates because the dephasing rates are set by noise power at specific high frequencies.  However, for dc gates, the dephasing rate due to $\de E_{ST}$  includes all quasistatic noise, and is about $10$ MHz \cite{gamblePRB12},  which can become the limiting relaxation rate for dc gates at large detuning, where detuning noise is strongly suppressed.}
Second, we can neglect the anisotropic relaxation term  $G_{XZ}$ which represent noise correlations in different directions on the Bloch sphere, as it is an order of magnitude smaller than the diagonal term $G_{XX}$  as shown in Fig.~\ref{fig:relax}a.
Note that  $G_{ZZ}$ and $G_{XZ}$ are  strongly suppressed because we  already tune the parameters to minimize charge dispersion ($\p B_Z/\p\e$).  
%Since the dominant term $S_XX$ give relaxation times of the order $10~\mu s$ (see Fig.~\ref{fig:relax}b), the approximations made here will no affect the gate fidelities 

While phonons cannot cause transitions between different spins, they could potentially cause decoherence as they couple to the difference in the charge distributions between singlet and triplet states in the doubly occupied dot with an interaction term that goes as ${H_{ep}\sim(a_{\rm ph}+a_{\rm ph}^\dag)(\ket{\cdot S}\<{\cdot S}|-\ket{\cdot T}\<{\cdot T}|)}$, where $a_{\rm ph}$ is the phonon destruction operator.\cite{huPRB11} 
However, averaging the fluctuations of the relative phase factor between $\ket{\cdot S}$ and $\ket{\cdot T}$ [see \eq{gamma1a}] over an incoherent thermal bath does not cause decay because the phonon density of states vanishes at low frequencies,\cite{huPRB11}  in contrast to the pure dephasing by $1/f$ detuning noise.
On the other hand, according to Ref.~\onlinecite{huPRB11}, when the phonon dissipative dynamics and coherent interaction with the qubit is fully taken into account, phonon relaxation can cause exponential decay of the singlet triplet coherence.
%\footnote{ with the average given by a quantum thermal expectation value, which would contain the usual factor of $\cosh(\hbar\w/\k_BT)$.  However, the noise power is proportional to the phonon density of states, which vanishes at low frequencies, so that pure dephasing $\propto S(0)$ does not occur.}
To model this effect, we will take the decay time ${T_{\rm ph}\simeq1~\mu}$s calculated in Ref.~\onlinecite{gamblePRB12} for silicon with a lateral electron confinement length of 40 nm. 
Since the phonon coupling has the same form as the charge noise coupling to $E_{ST}$, the phonon relaxation tensor is $F_{ij}$.  In particular, for the detunings of interest, only ${F_{ZZ}\simeq1}$ is relevant.
% cause $T_1$ charge relaxation of the hybrid qubit, 
% in principle, there is T_{1\rho} due to phonons at the Rabi freq, but should be small
Therefore, in the following, we will directly incorporate the decay due to the phonon dephasing in the off-diagonal density matrix elements by taking ${\rho_{01}(t)\to \rho_{01}(t)e^{-t/T_{\rm ph}}}$.

\subsection{Mitigating dephasing from low frequency noise\label{sec:mitlfnoise}}

%In \ref{sec:mitlfnoise}, we have shown that the high and low frequency noise spectrum of $(\de\w_R,\de\w_Z)$ gives a negligible contribution to dephasing of the driven qubit.

A key advantage of ac gate operations is that it is insensitive to low frequency transverse noise in $\de B_X$, resulting in  improved coherence during driven evolution compared to free induction decay.
Quantitative analysis of this effect is more convenient in the rotating frame, where the qubit simply undergoes free rotation about the $X$ axis, see \fig{fig:EffFields}d.
Applying the pure dephasing rate formula given in Eq.~\eqref{gamma1}, and noting that due to \eq{dbperp}, the transverse noise $\de\mbB_\perp'(t)$ has a shifted noise spectrum given by  \cite{smirnovPRB03}
%(see Appendix \ref{markov})
\ben
S'_{XX}(\w)= S'_{YY}(\w)=\frac{S_{XX}(\w+\w_Z)+S_{XX}(\w-\w_Z)}{4}\,,
\label{SXXrf}
\een
one finds the pure dephasing rate,
\ben
\frac{1}{T_{\phi}'}=\frac{S'_{XX}(0)}{2\hbar^2}=\frac{S_{XX}(\w_Z)}{4\hbar^2}\,.
\label{eq:Tphi}
\een
Thus, dephasing due to the low frequency transverse noise in $\de B_X$
% which is set by the quasistatic variance $\int_{\w_l}^{\w_R} d\w S_{XX}$
 is avoided.  The physical origin  for this effect can be understood by considering the RWA in the lab frame as illustrated in Fig.~\ref{fig:EffFields}c, where the qubit pseudospin precesses about the instantaneous, rotating transverse field 
\[\mbB_\perp(t)=(\hbar\w_R\cos\w_Zt+\de B_X(t))\mbf{e}_X+\hbar\w_R\sin\w_Zt\mbf{e}_Y\,.\]
The relative phase $\de\phi$ accumulated  due to fluctuations of the instantaneous rotation frequency is given by
\[\de\phi=\int dt'|\de\mbB_\perp(t')|/\hbar\simeq\int dt'\cos\w_Zt'\de B_X(t')\,,\]
which averages to zero for the low frequency components ($\w\ll\w_Z$) of $\de B_X$.
%Because the driving field $\mbB_{ac}$ rotates at a high frequency $\w_Z$, and the noise field $\de B_X$ has a fix direction along $X$, fluctuations of the Rabi frequency, corresponding to the magnitude $|\mbB_{ac}+\de B_X\mbf{e}_X|$, due to noise 
%Below, we will that transverse noise at the qubit  and Rabi frequency causes dephasing and relaxation of $X$ rotations.  
%Consistent with the Bloch-Redfield equations that we use in the following section.

\subsection{ac Bloch equations\label{sec:acBloch}}

In the section, we review the master equation in the Bloch-Redfield approximation in the rotating frame, valid when the Rabi frequency is much faster than the RF longitudinal relaxation rates, \cite{smirnovPRB03} which is satisfied in the parameter regime of interest in this work.
These equations include the dissipative qubit dynamics due to the noise term $\de\mbB_\perp'$ and  and $\de\w_Z$ in \eq{eq:HrfRWA}.
% at the Rabi frequency, which can cause $T_1$ relaxation in the rotating frame, as shown below.
% but its contribution will be negligible compared to that of $\de\mbB_\perp$.
%  ({This condition is given by ${S_{ZZ}(\w_R)\ll\w_R}$, see Eq.~\eqref{blochT}}).
The qubit master equation for  the pseudospin $\mbs=\<\psi|\bm{\s}|\psi\>/2$, where $\ket{\psi}$ the qubit state vector, is given by the Bloch equations  \cite{smirnovPRB03,jingPRA14,yanNATC13},
\begin{align}
\frac{d\mbs}{dt}&={\mbB\over\hbar}\times\mbs\nn
&-D_X(s_X-\bar{s}_X)\hatbf{e}_X-D_Ys_Y\hatbf{e}_Y-D_Zs_Z\hatbf{e}_Z
%\begin{pmatrix}\dot{s}_X'\\\dot{s}_Y'\\\dot{s}_Z'\end{pmatrix}
%=\begin{pmatrix}
%-D_X&0&0\\
%0&-D_Y&-\w_R\\
%0&\w_R&-D_Z
%\end{pmatrix}
%\begin{pmatrix}{s}'_X-\bar{s}'_X\\{s}'_Y\\{s}'_Z\end{pmatrix}
\label{eq:RFBloch}
\end{align}
where $\mbB$ is the effective field whose dominant term is $\hbar\w_R\mbf{e}_X$ (see \eq{eq:HrfRWA}), and
%\[\mbB=2\hbar\w_R\cos\w_Z t\hatbf{X}+u(t)\frac{\p B_Z}{\p u}\hatbf{Z}+\frac{u^2(t)}{2}\frac{\p^2\mbB}{\p u^2}+\ldots
%\,,\]
%and $\ldots$ are  higher order terms in u.  The 
the relaxation rates are given by
\begin{align}
D_X&={1\over T_{1\rho}}\nn
D_Y&={1\over T_\phi'}+{S_{ZZ}(\w_R)\over2\hbar^2}\nn
D_Z&={1\over T_\phi'}+{S_{YY}'(\w_R)\over2\hbar^2}\,.
\label{D}
\end{align}
The RF  $T_1$ and $T_2$ relaxation times are given by \cite{slichterPMR90}

\begin{align}
{1\over T_{1\rho}}&=\frac{S'_{YY}(\w_R)+S_{ZZ}(\w_R)}{2\hbar^2}\nn
\frac{1}{T_{2\rho}}&=\frac{D_Y'+D_Z'}{2}=\frac{1}{2T_{1\rho}}+\frac{1}{T_\phi'}\,,
\label{T2rf}
\end{align}
where the RF pure dephasing time $T_\phi'$ is given in Eq.~\eqref{eq:Tphi}.  Unlike the relaxation rates for dc rotation, these rates depend on both the Rabi and qubit frequency.

A plot of the relaxation times $(T_\phi',T_{1\rho},T_{2\rho})$ for both types of driving  as a function of detuning is shown in Fig.~\ref{fig:relax}c and an illustration of the associated relaxation processes is shown in Fig.~\ref{fig:relax}d.
For optimal parameters, ${G_{ZZ}\ll1}$ as shown in Fig.~\ref{fig:relax}a, so that $T_{2\rho}$, the decay time of ac $X$ rotations, is mainly due to noise power in the transverse components $S_{XX}\propto G_{XX}$, see \eq{GFtensors}, which scales quadratically with the detuning drive Rabi frequency $\w_R^2$, as mentioned in section \ref{sec:acgates} as a drawback of detuning drive.
This relaxation rate dominate the decay near the detuning sweet spot at ${\e_*\simeq20~\mu}$eV, where $T_{2\rho}\simeq10$ ns. 
As experimental evidence for this relaxation mechanism, we find that for the parameters of Ref.~\onlinecite{kimNATL15}, ${T_{2\rho}(\e_*)=5}$ ns, consistent with the short relaxation times found therein. 
%a and Fig.~\ref{fig:blochRates}b, respectively,
%Note that there is a slight anisotropy, $D_Z>D_Y> D_X$, and the dominant rate $D_Z$ is entirely due to transverse noise.  Similarly, 
On the other hand, beyond about $\e=200~\mu$eV, decoherence from detuning noise
% (except from low frequency noise)
is strongly suppressed, with all relaxation times above $1~\mu$s. 
At this point, phonon dephasing with $T_{\rm ph}=1~\mu$s  becomes the limiting relaxation mechanism.
% to $X$ rotation dephasing crosses over from detuning noise dominated to phonon dominated $T_{\rm ph}$ 
%While generally, all relaxation times depend on on the Rabi frequency, and hence depends on the type of driving,,
%Note that generally $T_{1,2\rho}$ and $D_X$ and $D_Z$ depends  but $T_\phi'$ and $D_X$ does not.  The dependence of $D_Z$ is also negligibly small, since $\w_R\ll\w_z$, see Eq.~\eqref{SXXrf}.
% the lab frame relaxation times given below Eq.~\eqref{T2rf}. 

The equilibrium pseudospin $\bar{\mbs}$ in general depends on the response of the noise bath to the driven qubit.  
Analytic expressions relating $\bar{\mbs}$ to the noise power spectrum are given in Ref.~\onlinecite{smirnovPRB03,yanNATC13}.
For the numerical simulations in section \ref{sec:fidelity},  we take the RF equilibrium condition ${\bar{\mbs}=-\hatbf{e}_X}$,  \cite{smirnovPRB03,jingPRA14} corresponding to equal populations of qubit states $|{\rm\pm}Z\>$ in the lab frame, which is valid in the low temperature limit  ${2k_BT\ll\hbar\w_{Z,R}}$, appropriate for typical temperatures ${T\sim0.1}$K of  experiments,  and the regime $ { T^{-1}_{\rho 1},T^{-1}_{\rho,2}\ll\w_R\ll\w_Z}$, appropriate near the optimal working point.

%\begin{table*}[t]
%\caption{Optimal gate fidelities obtained from (I) simulations of qubit dynamics using the three state Hamiltonian Eq.~\eqref{Hdc}, and numerically generated $1/f$ detuning noise. and (II) with the effective Hamiltonian}
%\begin{center}
%\begin{tabular}{ccccccc}
%method&drive type&(detuning,drive amplitude) $(\e,A_u)$ &gate time (ns) & ideal gate fidelity (\%)&actual gate fidelity&leakage probability(\%)\\
%\hline\hline
%I&$\D_{ac}$&(190,5)&1.7&99.9&99.5& 0.009\\
%II&$\D_{ac}$&(190,5)&1.7&99.9&99.5& 0.009\\
%I&$\e_{ac}$&(190,5)&1.7&99.9&99.5& 0.009\\
%II&$\e_{ac}$&(198,15)&1.7&99.9&99.1& 0.009\\
%\end{tabular}
%\end{center}
%\label{default}
%\end{table*}%

\subsection{Low frequency dephasing from noise in Rabi frequency\label{sec:lfdephase}}
\begin{figure}[t]
\begin{center}
\includegraphics[width=\linewidth]{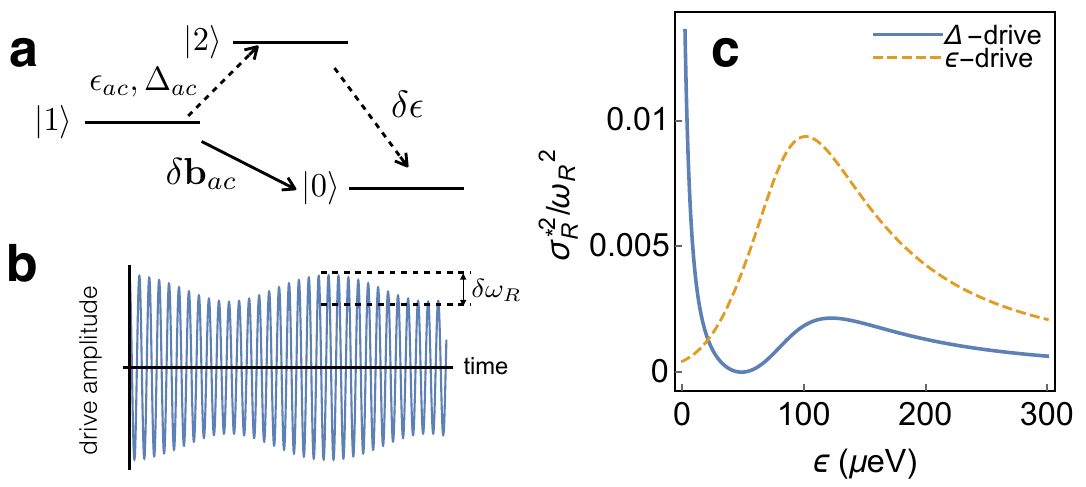}
\caption{(Color online) (a) Illustration of drive amplitude noise (solid line ) mediated by virtual  transitions (dashed lines) due to a combination of ac drive and detuning noise, described by second order term $\de\mbb_{ac}$ \eq{eq:bac1} in the effective Hamiltonian.
(b) Conceptual illustration of the fluctuations in the drive amplitude and associated Rabi frequency $\de\w_R$ due to low frequency detuning noise $\de\e$.
(c) For ${(\D^0_1,\D^0_2)=(30,30)~\mu}$eV, estimate of the mean square relative noise fluctuations in the Rabi frequency, $(\s^*_{R}/\w_R)^2$ [\eq{eq:sigmaR}]. Blue, solid  (gold, dashed) curve indicate detuning (coupling) drive.}
\label{fig:drabi}
\end{center}
\end{figure}

\begin{figure*}[t]
\begin{center}
\includegraphics[width=\linewidth]{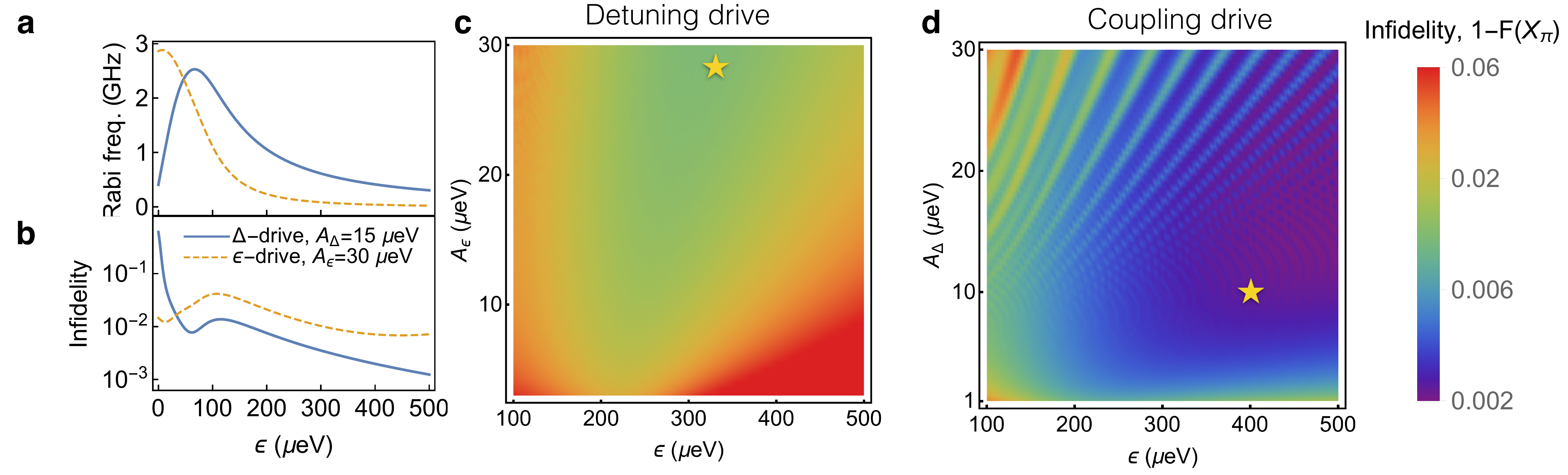}
\caption{(Color online) (a) Rabi frequency as a function of detuning $\e$, from \fig{fig:rabi}, shown here for comparison with infidelity.
(b) Estimate of the state infidelity of $X_\pi$ rotations \eqref{eq:infEst} based on the gate time, RF Bloch dephasing time $T_{2\rho}$ \eq{T2rf}, and the dephasing envelop $W_{\rm lf}(t_\pi)$ \eq{eq:Wlf1}.
(c and d)
Process infidelities for the ac $X_\pi$ gate $1-F(X_\pi)$, where the fidelity $F$ is defined above Eq.~\eqref{process}, as a function of detuning $\e$ and drive amplitudes $A_\e$ and $A_\D$, with colors scaled to Log ($1-F(X_\pi)$).
For detuning drive with amplitude $A_\e$ (c), an optimal point occurs at ${(\e,A_\e)=(316, 30)~\mu}$eV with $F(X_\pi)=99.1\%$.
For tunnel coupling  drive with amplitude $A_\D$ (d), an optimal point occurs at ${(\e,A_\D)=(400,10)~\mu}$eV with $F(X_\pi)=99.8\%$.
Both plots in (c) and (d) are computed with the static tunnel couplings $\D^0_1=\D^0_2=30~\mu$eV.}
% $\D_2=\D_2^0e^{-(\e-E_{ST})/\e_2}$,  and $\e_2=400~\mu$eV.}
\label{infidel}
\end{center}
\end{figure*}
In this section, we present our model for quasistatic dephasing due to noise in $\de\w_R$ and $\de\w_Z$ that is not captured by the Bloch relaxation rates \eq{T2rf} of section \ref{sec:acBloch}.
These noise terms cause low frequency  fluctuations of the instantaneous Rabi frequency $\de\w_R+\de\w_Z^2/2\w_R$ that is linear and quadratic in $\de\e$, respectively, \cite{yanNATC13,ithierPRB05,makhlinPRL04} see \fig{fig:EffFields}d and \eq{appeq:a12Z}.
An explicit calculation of the resulting low frequency decay envelope is given in appendix \ref{sec:lfnoise}.
Since we have already tuned the tunnel coupling to minimize ${\de\w_Z=\de\e\p\w_Z/\p\e}$ (${G_{ZZ}\ll1}$), the quadratic term in ${\de\w_Z^2/2\w_R=G_{ZZ}\de\e^2/2\hbar^2\w_R}$,  is negligibly small.\footnote{Dephasing due to high frequency noise in $\de\w_Z^2$ also gives a negligible contribution to gate infidelities, as discussed in appendix \ref{sec:hfnoise}.}
The main threat comes from low frequency noise in $\de\w_R$, which causes decay of RF $X$ rotations given by [cf.~Eq.~\eqref{Wlfexp}]
\ben
\begin{pmatrix}s_Y(t)\\ s_Z(t)\end{pmatrix}
=W_{\rm lf}(t)
\begin{pmatrix}s_Y(0)\\ s_Z(0)\end{pmatrix}\,,\,\, W_{\rm lf}=\exp\left(-\frac{t^2\s_R^2(t)}{2}\right)
\label{eq:Wlf1}
\een
where
\ben
\s_R(t)=\sqrt{\<\de\w_R^2\>_{\rm lf}}=\frac{\p\w_R}{\p\de\e}{\s_\e(t)}\,,
\label{eq:sigmaR}
\een
the angular brackets $\<\ldots\>_{\rm lf}$ denote an average over low frequency noise, 
and the quasistatic detuning noise variance is defined by [cf.~Eq.~\eqref{gamma1b}]
\ben\s^2_\e(t)={\<\de\e^2\>_{\rm lf}}={2\int_{\w_l}^{2\pi/t} {d\w\over2\pi} S_\e(\w )}={c_\e^2\over\pi}\ln\pfrac{2\pi}{\w_lt}\,.
\label{eq:lfdetune}\een
In the limit ${\w_lt\ll1}$ relevant to the qubit gate times ${t\sim1-10}$ ns, the logarithmic corrections dominate the decay.

 \eq{eq:lfdetune} will be included exactly in the simulations of section \ref{sec:fidelity}.  Here, we first give an estimate for the size of its effect on the gate infidelity, given approximately  by
\ben
1-W_{\rm lf}(t_g)\simeq\pfrac{\s^*_R}{\w_R}^2
%=\pfrac{\partial ^2b_X/\partial \epsilon \p u}{\partial b_X/\partial u }^2\<\de\e^2\>_{lf}
\label{eq:drabi}
\een
where ${t_g\propto1/\w_R}$ is the gate time, the effective detuning noise variance, ${\s_R^*=\s_\e^*{\p\w_R}/{\p\de\e}}$ is related to the dephasing time scale ${T_2^*(X)=\sqrt{2}\hbar/\s_R^*}$, where $\s_\e^*$ is given in Eq.~\eqref{sigstar}.
This infidelity estimate, for the optimal tunnel coupling parameters, is plotted in Fig.~\ref{fig:drabi}.  
It shows that for large detunings ${\e\simeq200~\mu}$eV, low frequency noise causes $<1\%$ infidelity for detuning drive and  $<0.5\%$ infidelity for coupling drive.  
We note this estimate is most likely an overestimate, since recent experiments indicate that the charge noise spectrum has a milder low frequency singularity than $1/f$.  
Specifically, Ref.~\onlinecite{dialPRL13} finds a spectrum of the form ${S_\e(\w)\propto 1/\w^\beta}$, with ${\beta=0.7}$.  
%Therefore, the infidelity estimate due to low frequency noise %We also note that the decay discussed here does not occur when the initial state lies on the $X$ axis, n the calculation of gate fidelities, detailed in appendix \ref{}, one of the {$T_{1\rho}$ immune to this dephasing} 

\section{Gate fidelities\label{sec:fidelity}}

In this section, we compute and optimize the quantum process fidelity of the ac $X_\pi$ (NOT) gate for both detuning and coupling driving.   
%using the numerical solution to the RF Bloch equations Eq.~\eqref{eq:RFBloch}.
We first consider the \emph{state} fidelity of an $X_\pi$ gate, defined as the probability of reaching the target state rotated by $X_\pi$ from the initial state $\ket{{\rm-}Z}$.
Since the total decay can be factorized  into the product of the low frequency decay envelop $W_{\rm lf}(t)$  Eq.~\eqref{eq:drabi} and the exponential decay envelop associated with the RF Bloch relaxation time $T_{2\rho}$ \cite{ithierPRB05}, the state fidelity is  given by
\ben
F_S(X_\pi)={e^{-t_\pi/T_{2\rho}}W_{\rm lf}(t_\pi)}
\label{eq:infEst}
\een
where ${t_\pi=\pi/\w_R}$ is the $X_\pi$ gate time.
The infidelity ${1-F_S(X_\pi)}$ as a function of detuning for both type of driving are plotted in Fig.~\ref{infidel}(b), which suggests that fidelities exceeding $99\%$ for both types of drive, and that coupling drive could yield fidelities near $99.9\%$.

We next describe our procedure for numerical simulations of the $X_\pi$ gate in the rotating frame.  For computing the process fidelity, it will be more convenient to use density matrix master equation,
\ben
\dot{{\rho}}=-\frac{i}{\hbar}[H_{RF},{\rho}]-D
\label{master}
\een
where $H_{RF}$ is the RF Hamiltonian \eq{eq:HRF}, \emph{without} taking the RWA, $D$ is the dissipator given by
\begin{align}
D&=D_X(s_X-\bar{s}_X){\s}_X+D_Ys_Y{\s}_Y+D_Zs_Z{\s}_Z\,,
\end{align} 
and ${\mbs(t)=\Tr(\rho(t)\bm{\s})/2}$ is the qubit pseudospin, [cf.~\eq{eq:RFBloch}].
% i.e.,$s_X=({\rm Re}\,{\rho}_{01})/2$, $s_Y=({\rm Im}\,{\rho}_{01})/2$, and $s_Z={({\rho}_{11}-{\rho}_{00})/2}$.
After obtaining the density matrix $\rho(t_\pi)$ after an $X_\pi$ gate by numerical integration of \eq{master}, we incorporate phonon-induced singlet triplet dephasing described in section \ref{sec:noisepower} and low frequency dephasing due to $\de\w_R$ described in section \ref{sec:mitlfnoise}, which cause the pseudospin components to decay as 
\begin{align}
s_X(t_\pi)&\to e^{-t_\pi/T_{\rm ph}}s_X(t_\pi)\nn
s_Y(t_\pi)&\to e^{-t_\pi/T_{\rm ph}}W_{\rm lf}(t_\pi)s_Y(t_\pi)\nn
s_Z(t_\pi)&\to W_{\rm lf}(t_\pi)s_Z(t_\pi)
\label{Ws}
\end{align}
where $t_\pi=\pi/\w_R$.  
Equivalently, we implement Eq.~\eqref{Ws} on the density matrix as 
\[\rho(t_\pi)\to e^{-t_\pi/T_{\rm ph}}(s_X{\s}_X+W_{\rm lf}(t_\pi)s_Y{\s}_Y)+W_{\rm lf}(t_\pi)s_Z{\s}_Z\,.\]
%\footnote{Note that the phonons only causes relaxation in the $X-Y$$, and thus do not directly affect the $X_\pi$ gate with initial condition $-Z$.} 

To find the optimal working point, we compute the quantum process fidelity of the $X_\pi$  gate as a function of detuning and ac drive amplitudes.
For a general quantum process $\mcal{E}$, the process fidelity is given by ${F(\mcal{E})=\Tr[\chi(\mcal{E})\chi(\mcal{E}_0)]}$, where $\chi$ is the process matrix defined by \cite{nielsenQC2010}
\ben
\mcal{E}({\rho})=\sum_{m,n}\hat{E}_m{\rho}\hat{E}_n\chi_{mn}\,.
\label{process}
\een
In our case, the process $\mcal{E}$ is the $X_\pi$ gate, $\mcal{E}(\rho)$ is the final density matrix after an $X_\pi$ gate computed from the simulation procedure described above, for an arbitrary initial density matrix $\rho$,  while the ideal process, denoted by $\mcal{E}_0$, is a perfect $X_\pi$ gate in the rotating frame  $\mcal{E}_0(\rho)=\s_X\rho \s_X$.
We follow the procedure for computing $\chi_{mn}$ given in Ref.~\onlinecite{nielsenQC2010},  with a basis set given by ${\hat{E}_m=\{1,\s_X,-i\s_Y,\s_Z\}}$. 

The $X_\pi$ gate infidelities for detuning and tunnel coupling drive are plotted as a function of detuning ($\e$) and drive amplitude ($A_u$) in Fig.~\ref{infidel}c and d.
In large regions of the parameter space $(\e,A_u)$, detuning drive fidelities exceed $99\%$ and tunnel coupling drive fidelity reach $99.8\%$.
These optimal regions are determined by the competition between i) gate speed, which favors small $\e$ and large $A_u$, ii) decoherence, which favors large $\e$, and iii) nonlinear effects, which favors small  $A_u$.   The  regions of high fidelity occur along the diagonal because  as $\e$ increases the drive amplitude needs to increase to maintain the same gate speed, which would otherwise decrease at fixed drive amplitude as shown in Fig.~\ref{infidel}a
The maximum fidelities are reached at ${(\e,A_\e)=(316, 30)~\mu}$eV with ${F(X_\pi)=99.1\%}$ for detuning drive and ${(\e,A_\D)=(400,10)~\mu}$eV and ${F(X_\pi)=99.8\%}$ for tunnel coupling driving. 

As a check on our decoherence model, we performed simulations of qubit dynamics using the three state Hamiltonian Eq.~\eqref{Hdc}, with $1/f$ detuning noise numerically generated by the procedure described in appendix \ref{sec:1overf}. 
The maximum fidelity for tunnel coupling driving computed with these simulations agree with the result based on Eq.~\eqref{master} and \eqref{Ws}.

\section{Discussions and conclusions}\label{sec:discussion}

The theoretical framework we have developed for analyzing and optimizing ac gate fidelities in this work are quite general, and can be applied to a generic noise power spectrum and to other semiconductor quantum dot qubits, such as the singlet triplet qubit.\cite{wongPRB15} 
While we considered silicon, the three-electron double dot hybrid qubit can be implemented in other materials such as Germanium or Gallium arsenide.\cite{cao15}  Our theory will be applicable in these materials provided that the qubit and noise parameters are adjusted appropriately.  
For example, the qubit frequency will be set by the single dot singlet triplet splitting in these materials.  The phonon relaxation rates depends on the character of phonon excitations, attenuation rates, and the lateral dot radius, which is set by the transverse effective mass and gate-defined confinement potential.\cite{huPRB11,{gamblePRB12}}

While we were focused in this work on how charge noise affect qubit fidelities,  the decoherence model we have developed can be used to turn the question around, to probe the charge noise spectrum by measuring qubit dynamics.\cite{yanNATC13}
The information thus gained about the nature of the environment causing the charge noise can be helpful in developing experimental and fabrication techniques to reduce it.

In summary, we have systematically analyzed gate fidelities of the ac driven quantum dot hybrid qubit, including decoherence due to $1/f$ charge noise, determined the optimal parameter regime for the tunnel coupling, detuning, and ac drive strengths, and showed that gate fidelities up to 99.8\% can be achieved by driving the tunnel coupling.
The fidelities computed in this work are exponentially sensitive to the $1/f$ detuning noise parameter $(c_\e)$, so that we expect even modest reduction in charge noise could result in significant further improvements in qubit fidelity. 
The ac driven single qubit operations studied in this work are a crucial part of two-qubit gate sequences such as the one proposed in Ref.~\onlinecite{mehlCM15}, 
and should thus enable a high fidelity universal gate set for quantum dot hybrid qubit.

\textit{acknowlegements}--We thank Mark Friesen and S. N. Coppersmith for guidance and stimulating discussions. This work was supported by the Intelligence Community Postdoctoral Research Fellowship Program.

\appendix

\section{Derivation of effective Hamiltonian\label{app:canon}}
\begin{figure}[b]
\begin{center}
\includegraphics[width=.5\linewidth]{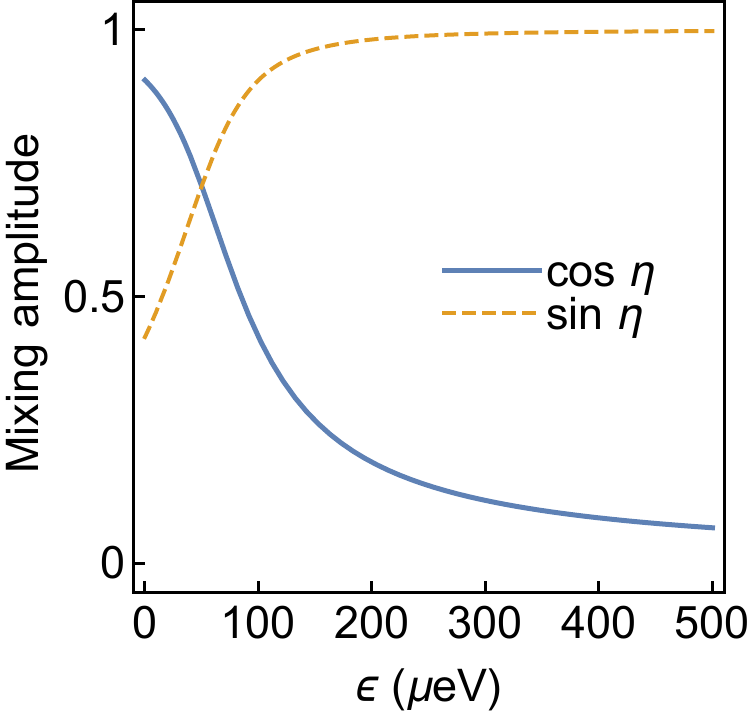}
\caption{(Color online) The mixing amplitudes ${\cos\eta=\bra{S\rm\cdot}\ket{1}}$ and  ${\sin\eta=\bra{{\rm\cdot} T}\ket{1}}$ in Eq.~\eqref{eq:mixing} as a function of detuning $\e$.}
\label{fig:mixing}
\end{center}
\end{figure}

In this appendix, we derive the effective Hamiltonian for the qubit, following the standard procedure,\cite{foldyPR50,winklerS0Book03} including modifications due to dynamics.
We first identify the leakage state that is separated from qubit subspace by a finite gap at all detunings. To this end, we first diagonalize  the ${\{|\cdot T\> ,|S \cdot \>\}}$ subspace, where the tunnel coupling $\D_2$ causes hybridization of the $S(2,1)$ and $T(1,2)$ states.   The charge and spin hybridized eigenstates are 
\[\begin{pmatrix}\ket{1}\\|\til{L}\>\end{pmatrix}=
\begin{pmatrix}
 \sin\eta_0  &    \cos\eta_0 \\
 \cos\eta_0  & - \sin\eta_0
\end{pmatrix}
\begin{pmatrix}{|\cdot T\>}\\{|S \cdot \>}\end{pmatrix}\]
where
\ben
\begin{pmatrix}\cos\eta_0\\\sin\eta_0\end{pmatrix}
=\frac{1}{\sqrt{2 \til{E}_{12}}}\begin{pmatrix}\sqrt{\til{E}_{12}+E_{{\rm ST}}-\epsilon_0}\\\sqrt{\til{E}_{12}-E_{{\rm ST}}+\epsilon_0}\end{pmatrix}\,,
\label{eq:mixing}
\een
 $\eta_0$ is the $\e_0$ dependent mixing angle, and
\[\til{E}_{12}=\sqrt{4 \til{\D}_2^2(\e_0)+(E_{{\rm ST}}-\epsilon_0)^2}\,,\]
is the energy splitting, and we define the detuning dependent tunnel coupling $\til{\D}_2(\e_0)=\D_2^0e^{-(\e_0-E_{ST})/\e_2}$.  The mixing amplitudes $(\cos\eta,\sin\eta)$ as functions of detuning are plotted in Fig.~\ref{fig:mixing}.

With this change of basis at $\e_0$, defined by the following transformation
\ben
U_\eta(\e_0)=\begin{pmatrix}
 1 & 0  &0   \\
 0 &   \sin\eta_0  &    \cos\eta_0 \\
 0 &   \cos\eta_0  & - \sin\eta_0
\end{pmatrix}
\label{eq:U0}
\een
the  Hamiltonian becomes
\begin{widetext}
\begin{align}
H_\eta(\e,\D_1,\D_2)=U_\eta^\dag H_0U_\eta
=\begin{pmatrix}
 -\frac{\e(t)}{2} & \D_1(t) \cos \eta_0 & -\D_1(t) \sin\eta_0 \\
 \D_1(t) \cos \eta_0 & \frac{ \e(t)}{2} \cos 2\eta_0-\D_2(t) \sin 2\eta_0 & \frac{E_{ST}-\epsilon(t)}{2}\sin 2\eta_0-\D_2(t) \cos 2\eta_0 \\
 -\D_1(t) \sin\eta_0 & \frac{E_{ST}-\epsilon(t)}{2} \sin 2\eta_0-\D_2(t) \cos 2\eta_0 & \D_2(t) \sin 2\eta_0-\frac{ \e(t)}{2} \cos 2\eta_0 \\
\end{pmatrix}\,,
\label{H1}
\end{align}
where, turning on both detuning and coupling drive,
\[\e(t)=\e_0+\e_{ac}(t)\,,\quad  \D_1(t)=\D_1^0+\D_1^{ac}(t)\,,\quad \D_2(t)=\til{\D}_2(\e_0)+\D_2^{ac}(t)\,.\]
%where $\D_{1ac}=\D_{ac}(t)$ and  $\D_{2ac}=\D_{ac}(t)e^{-(\e(t)-E_{ST})/\e_2}$.
Writing ${H_\eta=H^\eta_{dc}+H^\eta_{ac}}$, the dc Hamiltonian is given by 
\ben
H^\eta_{dc}(\e_0,\D_{1}^0,{\D}_2^0)=\begin{pmatrix}
 -\frac{\epsilon_0 }{2} & \D_1^0\cos\eta_0&-\D_1^0\sin\eta_0 \\
 \D_1^0\cos\eta_0& \frac{1}{2} \left(E_{ST}-\til{E}_{12}^0\right) & 0 \\
-\D_1^0\sin\eta_0  & 0 & \frac{1}{2} \left(E_{ST}+\til{E}_{12}^0\right) \\
\end{pmatrix}\,,
\label{Hdc}
\een
and the ac part is given by
\[H^\eta_{ac}(t)=H_\eta(\e(t);\D_1(t),\D_2(t))-H^\eta_{dc}
=\begin{pmatrix}
 -\e_{ac}/2& \D_1^{ac} \cos \eta_0 & - \D_1^{ac}  \sin\eta_0 \\
  \D_1^{ac}  \cos \eta_0 & (\e_{ac}/2) \cos 2\eta_0- \D_2^{ac}  \sin 2\eta_0 & -(\e_{ac}/2)\sin 2\eta_0- \D_2^{ac}  \cos 2\eta_0 \\
 - \D_1^{ac}  \sin\eta_0 &- (\e_{ac}/2) \sin 2\eta_0- \D_2^{ac}  \cos 2\eta_0 &  \D_2^{ac}  \sin 2\eta_0-(\e_{ac}/2) \cos 2\eta_0 \\
\end{pmatrix}\,.\]
\end{widetext}
Detuning noise can be included simply by taking ${\e_{ac}\to\e_{ac}+\de\e}$. So far, no approximations have been taken. { Note that $H_\eta$ has different dependences on $\e_{ac}$ and $\e_0$, due to the $\e_0$-dependent transformation \eq{eq:U0}.}

Next, we perform a canonical transformation to derive the effective Hamiltonian for the ac driven qubit.  This transformation is a perturbative expansion organized as follows.
We first separate $H_\eta$ into a diagonal ${H_d={\rm  diag} (H)}$ and an off-diagonal ${V^{(0)}=H-H_d}$ part, which is considered to be a perturbation, so that ${H_\eta=H_d+V^{(0)}}$.  The off-diagonal part is further separated into ${V^{(0)}=V_Q+V_L}$, where  $V_Q$ is off-diagonal within qubit subspace, $V_L$ is  off-diagonal between qubit and leakage space.  $V_Q$ is justified  as a perturbation even for large $\D_1$ because in the far detuned regime ${\cos\eta_0\ll1}$ (see Fig.~\ref{fig:mixing}).
%Note that we include the time dependent driving terms on the diagonal in $H_d$, and that 
We then perform the canonical transformation 
\[|\psi\>=U_S|\til{\psi}\>\,, \quad U_S=e^{S}\,,\]
where $S$ is anti-Hermitian ($S^\dag=-S$) and a is purely leakage term. 
The transformed Hamiltonian, up to $O(S^3)$ terms, is given by
\begin{align}
\til{H}&\equiv U_S^\dag HU_{S}-i\hbar U_S^\dag\p_tU_S\nn
&=H-i\hbar\dot{S}+[S,H+{i\hbar\over2}\dot{S}]+{1\over2\!}[S,[S,H+{i\hbar\over3}\dot{S}]]\nn
%&=H_d+V+\dot{S}+[H_d,S]+[V,S]+{i\hbar\over2}[S,\dot{S}]+{1\over2}[[(H_d,+V)S],S]\,.
&=H_d+V_Q+[V_L,S]+{1\over2}[[H_d,S],S]+{i\hbar\over2}[S,\dot{S}]\nn
&+{V_L+[H_d+V_Q,S]+i\hbar\dot{S}+{1\over2}[[V,S],S]}
\label{eq:tilH}
\end{align}
The leakage terms, given in the last line of \eq{eq:tilH}, are eliminated perturbatively in the parameters ${\al_n=V^{(0)}/\D E_{nl}}$ and ${\beta_n=\hbar\w_Z/\D E_{nl}}$, where ${\D E_{nl}=E^\eta_{n}-E^\eta_{l}}$,  $E^\eta_{n,l}$ are the diagonal elements of $H^\eta$, $n (l)$ labels states in the qubit (leakage) subspace.  For static  Hamiltonians, $\al_n$ is sufficient to parametrize the perturbation theory. 
In the dynamic case, the additional expansion parameter $\beta$ comes from the dynamics described by $\dot{S}$. We will denote the $k$th order of the perturbative series as the term of power $\al_m^p\be_n^q$, with ${k=p+q}$.  
The decoupling procedure is iterative: \cite{foldyPR50} $S=\sum_{k=1}^\infty S^{(k)}$ is a perturbative series with each term of order $k$, chosen to eliminate leakage terms of order $\al_n^{k-1}$, and produces corrections in the effective Hamiltonian as well as leakage terms of order $k$.  

In the leading order,  $S=S^{(1)}$ is chosen to satisfy
\[[H_d,S^{(1)}]=-V_L\,,\quad S^{(1)}=-\frac{V^{(0)}}{\D E_{nl}}\,,\]
{The leakage term to the next order, generated by $S^{(1)}$ is given by
\[V^{(1)}=[V_Q,S^{(1)}]+{i\hbar\p_t S^{(1)}}\,,\] which can be eliminated with $S^{(2)}=-{V^{(1)}}/{\D E_{nl}}$, resulting in terms of order $\al_m\al_n$ and $\al_m\beta_n$ in the effective Hamiltonian.}

We will keep the perturbation theory to leading order, where the nonzero matrix elements of $S^{(1)}=S_{dc}+S_{ac}(t)$, $S_{dc}$ being the static part and $S_{ac}$ the time dependent part,   are given by
\begin{align}
\<{2}|S_{dc}\ket{0}&=\frac{2\D_1^0\sin\eta_0}{\D E_{02}}\,,\quad 
\<2|S_{ac}(t)\ket{0}=\frac{\D_1^{ac}  \sin\eta_0}{\D E_{02}}\nn
\<2|S_{ac}(t)\ket{1}&=-\frac{(\bar{\epsilon}/2)\sin 2\eta_0-\D_2^{ac} \cos 2\eta_0 }{\D E_{12}}
\end{align}
The  effective Hamiltonian is then given by
\ben
h=P_\eta\til{H} P_\eta=H_d+V_Q+{1\over2}[V_L,S^{(1)}]\,,
\label{Heff}
\een
where $P_\eta=\ket{\eta_1}\<\eta_1|+\ket{\eta_2}\<\eta_2|$ is the projection operator onto the qubit subpsace, defined as the lowest two energy eigenstates $\ket{\eta_{0,1}}$ of ${H}_\eta$ at $t=0$.  The matrix elements are given by
\begin{widetext}
\begin{align}
h_{00}(t)&=-\frac{\e(t)}{2}
-\frac{2 \Delta _1^2(t) {\sin^2\eta_0 }}{{\cos 2\eta_0 } (E_{{ST}}-\e(t) )+2 \Delta _2(t) {\sin 2\eta_0 }+E_{{ST}}+\e(t) }\nn
h_{01}(t)&=h_{10}=\frac{1}{4} \Delta _1(t) \left({\cos\eta_0 }+\frac{{\cos\eta_0 } (E_{{ST}}-\e(t) )+2 \Delta _2(t) {\sin\eta_0 }}{{\cos2\eta_0 } (E_{{ST}}-\e(t) )+2 \Delta _2(t) {\sin2\eta_0 }}+\frac{4 {\cos\eta_0 } E_{{ST}}+4 \Delta _2(t) {\sin\eta_0 }}{{\cos2\eta_0 } (E_{{ST}}-\e(t) )+2 \Delta _2(t) {\sin2\eta_0 }+E_{{ST}}+\e(t) }\right)\nn
h_{11}(t)&=\frac{1}{4}\left(2 E_{{ST}}-\frac{ \Delta _2^2(t)+(E_{{ST}}-\e(t))^2}{{\cos2\eta_0 } (E_{{ST}}-\e(t) )+2 \Delta _2(t) {\sin2\eta_0 }}+{\cos2\eta_0 } (\e(t) -E_{{ST}})-2 \Delta _2(t) {\sin2\eta_0}\right)\,,
\label{appeq:h}
\end{align}
\end{widetext}
We note here that the difference in the $\e_0$ and $\e_{ac}$ dependence comes from the implicit $\e_0$ dependence in $\eta_0$, and in particular the difference in the respective derivatives is given by 
\ben
\frac{\p h}{\p\e_0}-\frac{\p h}{\p\e_{ac}}=\frac{\p\eta}{\p\e_0}\frac{\p}{\p\eta}\,,
\label{eq:deps}
\een
where $\eta$ is the mixing angle defined in \eq{eq:mixing}.  This difference is important at low detunings, where $\eta$ has strong detuning dependence, as shown in \fig{fig:mixing}. 
 
The static effective Hamiltonian that follows by taking $\e(t)\to\e_0$ is  given by 
\ben
H_{\rm dc}=\begin{pmatrix}
-{\e_0/2}-\frac{2\D_1^2\sin^2\eta_0}{\til{E}_{12}+E_{ST}+\epsilon}&\D_1\cos\eta_0\\
\D_1\cos\eta_0&E_{{\rm ST}}-\til{E}_{12}/2\\
\end{pmatrix}
\label{eq:Heffdc}
\een
and the qubit  basis is 
\ben
|{Q}_n\>\simeq(1-S_{dc})U_\eta|q_n\>
\label{eq:basis}
\een
where $|q_n\>$ labels the basis states $\{|\cdot S\>,|S\cdot\>,|\cdot T\>\}$.  

\section{longitudinal driving\label{LdriveApp}}

\begin{figure}[t]
\begin{center}
\includegraphics[width=\linewidth]{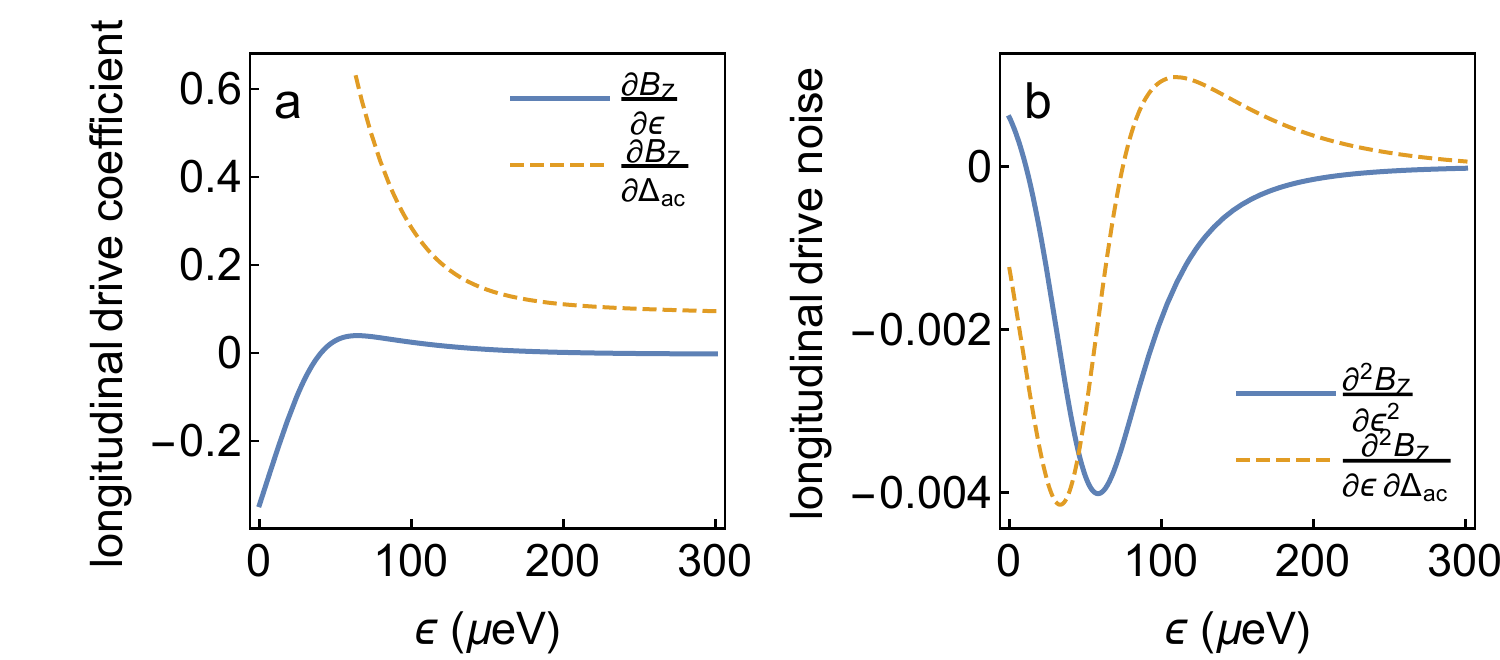}
\caption{(Color online) As a function of detuning $\e$:
(a)  The longitudinal drive coefficient for detuning and tunnel coupling drive, $\p B_Z/\p g$.
(b) The coefficient of the noise fluctuation the longitudinal drive amplitude ${\p^2 B_Z}/{\p\e\p u}$ for both types of driving.}
\label{fig:longdrive}
\end{center}
\end{figure}

The longitudinal driving field  in Eq.~\eqref{bac} is given by
\ben
B^{(1)}_{acZ}(t)=A_u\cos\w_Z t\left(\frac{\p B_Z}{\p u}+\de\e\frac{\p^2B_Z}{\p(\de\e)\p u}\right)\,.
\label{eq:longdrive}
\een
Consider first the dominant contribution in Eq.~\eqref{eq:longdrive} proportional to ${\p B_Z}/{\p u}$, plotted in Fig.~\ref{fig:longdrive}a  as a function of $\e$ for both types of driving,
This term can cause modulations and changes in Rabi frequency when ${\w_R\simeq\w_Z}$.  However,  near the optimal working point of this qubit, $\w_R\ll\w_Z$, so that these effects are strongly suppressed.  
For detuning driving,  this term, given by ${\p B_Z/\p\e=\p E_{10}/\p\e}$, was already minimized  in \ref{sec:chargedisp}, and is essentially zero,
As shown in Fig.~\ref{fig:longdrive}(a).
For coupling driving,  ${\p B_Z/\p\D_{ac}\sim0.1}$.
The second term in  Eq.~\eqref{eq:longdrive} represent noise in the drive amplitude that has the same origin as $\de\w_R$, the noise in the Rabi frequency.  In principle, this noise term should be added to $\de B_Z$, which could cause $T_{1\rho}$ relaxation.  However, it is strongly suppressed because: 
i) Due to the prefactor $\cos\w_Z t$, the noise spectrum is shifted similarly to $\de B_X$.  In particular, the noise power that would contribute to $T_{1\rho}$ occurs at $S_\e(\w_Z\pm\w_R)$, which is reduced by an order of magnitude from $S_\e(\w_R)$, and 
ii) The drive noise coefficient $\p^2B_Z/\p u\p(\de\e)$, plotted in Fig.~\ref{fig:longdrive}b, is of order $10^{-3}$.

\section{$1/f$ noise\label{app:noise}}

In this section, we summarize properties of $1/f$ noise.  

\subsection{Noise correlations}

A parameter $V$ with Gaussian noise $\de V$ is completely characterized by its autocorrelation function, defined by
\begin{align}
S_{VV}(t-t')=\<\de V(t)\de V(t')\>=\int_{-\infty}^\infty {d\w\over2\pi}\,e^{i\w(t-t')}S_V(\w)\,,\nonumber
\label{SV}
\end{align}
where ${S_V(\w)=\<|\de V(\w)^2|\>}$ is the noise power density.  The variance isgiven by
\ben
\s_V^2=\<\de V^2(t)\>=S_V(t=t')=\int_{-\infty}^\infty {d\w\over2\pi}\,S_V(\w)\,,
\label{sigma}
\een
and the correlation time is defined by
\ben
\tau_V=\frac{1}{\s_V^2}\int_0^\infty dt' S(t')\,.
\label{tauc}
\een
For a noise source with exponential correlations, ${S(t)=\s_V^2 e^{-t/\tau_V}}$. 
% \mf{ornstein Uhlenbeck}.
%  The Bloch equations are valid when $\tau_c\ll$ relaxation times.  
For $1/f$ noise, a high ($\w_h$) and low ($\w_l$) frequency cutoff is necessary to make $\tau_V$ and $\s_V$ finite.  The low frequency cutoff sets the correlation time, $\tau_V=2\pi/\w_l$.

\subsection{Experimental fit of detuning noise strength\label{sec:T1fit}}
We fit the single parameter $c_\e$ in the noise spectrum using the $T_1$ relaxation time for decay from $|1\>\to|0\>$,  measured to be 7 ns at the sweet spot in Ref.~\onlinecite{kimNATL15}.  A similar time scale was measured in Ref.~\onlinecite{peterssonPRL10}, which attributes it to charge noise.
We thus fit the detuning noise strength to the $T_1$ relaxation time using the  Bloch formula
\[\frac{1}{T_1}=\frac{S_{XX}(\w_Z)}{2\hbar^2}=\frac{G_{XX}S_\e(\w_Z)}{2\hbar^2}\,,\]
from which we find 
\ben
c_\e=\sqrt{\frac{2\hbar^2 \w_Z(\e_*)}{G_{XX}(\e_*)T_1(\e_*)}}=2.38~\mu {\rm eV}^2\,,
\label{eq:ce}
\een
where the detuning sweet spot for parameters in \cite{kimNATL15} is located at $\e_*\simeq20~\mu$eV.

\subsection{Pure dephasing rates due to {$1/f$ noise}\label{sec:dephase1}}
 
In this section, we review the relevant formulae for pure dephasing rates for $1/f$ noise.
Consider two quantum states denoted  generically as $|\pm\>$, with an  energy splitting ${E=E_+-E_-}$ and noise fluctuations ${\de E=\de E_+-\de E_-}$, which  for definiteness we assume in the section to come from detuning noise $\de\e$.
Pure dephasing refers to the decay in the off-diagonal element of the density matrix ${\rho_{+-}=\<+|\rho|-\>=e^{-i\phi}\<e^{-i\de \phi}\>/2}$, where ${\phi=(E_+-E_-)t/\hbar}$ is the relative phase and 
%and $\de\phi=\de Et/\hbar$ is the accumulated phase difference due to noise.
% which represents the relative phase factor  between  
\ben\de \phi(t)=\int_0^t dt'\de E(t')/\hbar\,,\label{eq:dphi}\een
is the accumulated phase due to noise fluctuations, which typically causes decay  when averaged over noise realizations.  We keep to quadratic order in detuning noise, so the energy fluctuation is given by
\begin{align}
{\de E(t)}&=a_1\de \e(t)+{a_2\over2}\de\e(t)^2\,,\nn 
(a_1,a_2)&=\left(\frac{\p E}{\p (\de\e)},\frac{\p^2 E}{\p(\de\e)^2}\right)\,.
\label{eq:dE}
\end{align}
The decay envelope is given by 
\ben
W(t)\equiv{\<e^{-i\de\phi(t)}\>}=\exp\left[-\G^{(1)}_\phi(t)-\G^{(2)}_\phi(t)\right]\,,
\label{W}
\een
where $\G^{(1)}_\phi(t)$ and $\G^{(1)}_\phi(t)$ are due to noise averaging over the linear  $(\de\e)$  and quadratic $(\de\e^2)$ term, respectively. 
The linear term can be expressed in  a well known,  closed form
\begin{align}
\G^{(1)}_\phi(t)&=\onehalf\int_0^t\int_0^t dt_1dt_2\<\de E(t_1)\de E(t_2)\>\label{gamma1a}\\
&={a_1^2\over2}\int_0^t\int_0^t dt_1dt_2\<\de \e(t_1)\de \e(t_2)\>\nn
&=\frac{a_1^2t^2}{2\hbar^2}\int_{-\infty}^\infty {d\w\over2\pi} 
S_{\e}(\w){\rm sinc}^2(\w t/2)\,,
\label{gamma1}
\end{align}
where ${{\rm sinc}(x)={\sin(x)}/{x}}$.  Note that due to the sinc function, the integral is dominated by the quasistatic part of the spectrum  (${\w<1/t}$).
Physically, this stems from the fact that the phase accumulation $\de\phi$ coming from  noise at frequencies higher than $1/t$  tends to time-average to zero in \eq{eq:dphi}.

The decay exponent due to quadratic fluctuations cannot in general be expressed in closed form, but it can be expressed in terms of a functional determinant
 \begin{align}
\G_\phi^{(2)}&=-\ln{[\det(1+i{a_2S/\hbar})]^{-1/2}}\nn
&=\onehalf\tr\ln(1+i{a_2S/\hbar})\nn
&=\onehalf\sum_n{1\over n}\tr[(-i{a_2S/\hbar})^n]
%&=\onehalf\left({-ia_2\over\hbar}\tr S +\frac{a_2^2}{2\hbar^2}\tr S^2+\ldots\right)\,.\notag
\label{gamma2}
\end{align}
where the multiplicaton of noise correlation functions denotes integration over functional kernals
 \begin{align*}
 \tr S&=\int dt' S(t',t')\,,\nn
\tr\, S^2&=\int dt'dt'' S(t',t'')S(t'',t')\,.
 \end{align*}
The series expression in Eq.~\eqref{gamma2} was previously derived in Ref.~[\onlinecite{makhlinPRL04}], and it can be simply generalized to include  dynamical decoupling sequences such as spin echo or CPMG \cite{cywinskiPRB08}.

At a fixed gate time $t_g$, the summation  Eq.~\eqref{gamma2}  can be evaluated for high and low frequency ranges relative to $1/t_g$ \cite{makhlinPRL04}.
In contrast to the linear term Eq.~\eqref{gamma1}, \emph{both} high and low frequency ranges contribute at all times, but the decay is dominated at short (long) times by the low (high) frequency contribution.    Here, the short (long) time regime is defined relative to the time scale, \cite{makhlinPRL04,ithierPRB05}
\ben
\tau_2=\frac{\hbar}{a_2c_\e^2}\,.
\label{eq:tau2}
\een 
%and the decay behavior is near-Gaussian, algebraic, and exponential for ${t\ll\tau_2}$, ${t\sim\tau_2}$, and ${t\gg\tau_2}$, respectively 

For the optimal parameters considered in this work, we will be in the short time regime ${t\ll\tau_2}$, and the quadratic noise terms are strongly suppressed.  However, these higher order effects may be important in the experiments that operate away from these optimal parameters, for example, in Ref.~\onlinecite{kimNATL15,kimCM15}.
% We encounter the long time regime when we fit to data from previous experiments in  \ref{values}.
%We will apply the dephasing formulas Eq.~\eqref{gamma1} and Eq.~\eqref{gamma2} to dc $Z$ rotations in the lab frame and ac $X$ rotations in the rotating  frame. 
\subsubsection{Linear and quadratic noise couplings  \label{dephaseR}}
In this section, we determine the  linear ($a_{1}$) and quadratic ($a_{2}$) couplings to detuning noise [\eq{eq:dE}] coming from the leading order expansion in $\de\e$ of the noise field given in  \eq{db}.  These coefficients can be used in the formulae given in appendix \ref{sec:lfnoise} and \ref{sec:hfnoise} to compute  dephasing envelops for dc $Z$ rotations in the lab frame and ac $X$ rotations in the rotating frame as described

In the lab frame, the fluctuations in the qubit frequency  (dc $Z$ rotation frequency)  is given to quadratic order by
\[{\de E}=\sqrt{(B_Z+\de B_Z)^2+\de B_X^2}-B_Z=\de B_Z+\frac{\de B_X^2}{2B_Z}\,,\]   
where $B_Z=\hbar\w_Z$,  and we define ${\de f\equiv\de\e\p f/\p(\de\e)}$ for a generic function of detuning $f(\de\e)$.    It follows that
\ben
a_1(Z)=\frac{\p B_Z}{\p(\de\e)}\,,\quad a_2(Z)=\frac{(\p B_X/\p(\de\e))^2}{B_Z}\,.
\label{appeq:a12Z}
\een
For ac $X$-rotations, in the RF and RWA, the fluctuations of the Rabi frequency is given by
\ben
{\de E\over\hbar}=\sqrt{(\w_R+\de\w_R)^2+\de\w_Z^2}-\w_R\approx\de\w_R+\frac{\de\w_Z^2}{2\w_R}
\label{eq:dEac}
\een
so that 
\begin{align}
a_1(X)&=\hbar\frac{\de\w_R}{\de\e}={A_u\over2}\frac{\p^2B_X}{\partial(\de\e)\p u}\,,\nn
a_2(X)&={\hbar\over\w_R}\frac{\de\w_Z^2}{\de\e^2}=\frac{(\p B_Z/\p(\de\e))^2}{\hbar\w_R}\,,
\label{eq:a12X}
\end{align}
where  ${u=\e_{ac}}$ or $\D_{ac}$.

\subsubsection{Dephasing due to low frequency noise\label{sec:lfnoise}}
Next, we consider the quasistatic contributions to dephasing for linear Eq.~\eqref{gamma1} and quadratic Eq.~\eqref{gamma2}  terms in turn.  
Due to $1/f$ singularity,  the decay exponent in Eq.~\eqref{gamma1} is dominated by low frequency noise at $\w\ll1/t$, so that we can take sinc${(\w t/2)\simeq1}$ in the integrand.
This yields  a Gaussian-like decay given by 
\begin{align}
\G^{(1)}_\phi(t)&=\frac{t^2}{2\hbar^2}a_1^2\s_\e'^2(t)\,,
\label{gamma1a}\\
\s'_\e(t)&=\sqrt{2\int_{\w_l}^{2\pi/t} {d\w\over2\pi} S_\e(\w )}=c_\e\sqrt{{1\over\pi}\ln\pfrac{\tau_m}{t}}\,.
\label{gamma1b}
\end{align}
When the low frequency term \eqref{gamma1a} dominates the decay \eqref{W}, the dephasing time scale is defined by ${1=\G^{(1)}_\phi(T_2^*)}$, and   is given approximately by \cite{makhlinPRL04} ${T_2^*\approx{\tau_1}/{\sqrt{\ln (\tau_m/\tau_1)}}}$, where  ${\tau_1={\sqrt{2}\hbar}/{a_1c_\e}}$.
If we define a variance by the time scale ${T_2^*=\sqrt{2}\hbar/a_1\s_\e^*}$, then 
\begin{align}
 \s^*_\e&= c_\e\sqrt{(1/\pi)\ln(\tau_m/\tau_1)}\simeq c_\e\sqrt{(1/\pi)\ln(c_\e \tau_m/h)}\nn
 &=5.5~\mu \rm{eV},\label{sigmastar} \end{align}
For the ac gates times of ${t_g\sim1}$ ns, the ${\s_\e'(t_g)=5.9~\mu}$eV, which is approximately equal to $\s^*_\e$.  
%However, we note that for faster rotations such as the $Z$ gates, the logarithmic time dependence in $\s_\e'(t)$ can dominate the decay behavior.

The dephasing exponent due to the quadratic term for a $1/f$ noise spectrum Eq.~\eqref{gamma2} is evaluated in detail in Ref.~\onlinecite{makhlinPRL04}, where it is shown that the low frequency contribution to  $\G_\phi^{(2)}$
%at times up to $t\sim\tau_2$
 is given by
\begin{align}
\G^{(2)}_\phi(t)
=-\onehalf\ln(1+ia_2t\s_\e^{\prime2}(t)/\hbar)\,.
%=-\onehalf\ln\sqrt{1+[a_2t\s_\e^{\prime2}(t)]^2}-i\tan^{-1}[a_2t\s_\e^{\prime2}(t)]
\label{gamma2a}
\end{align}
This  is the dominate contribution for times up to ${t\sim\tau_2/\ln(\tau_m/\tau_2)}$.
For short times ${t\ll\tau_2}$, it gives a Gaussian-like decay similar to Eq.~\eqref{gamma1b}.
%where the decay becomes algebraic  $\sim1/\sqrt{t}$ 

To summarize, the low frequency (quasistatic) contributions to the total decay envelope Eq.~\eqref{W} is given by
\ben
W_{\rm lf}(t)=\frac{\exp\left[{-(a_1t\s_\e'(t)/\sqrt{2}\hbar)^2}\right]}{\sqrt{1+ia_2t\s_\e^{\prime2}(t)/\hbar}}
\label{Wlf}
\een
For $t\ll\tau_2$,   the decay envelop is given by
\ben
|W_{\rm lf}(t)|=\exp\left[-(\g_\phi(t) t)^2\right]\,,
\label{Wlfexp}
\een
where the decay rate is given by
\ben
\gamma _{\phi}(t)={1\over\hbar}\sqrt{\pfrac{a_1\s_\e'(t)}{\sqrt{2}}^2+\pfrac{a_2\s_\e^{\prime2}(t)}{2}^2}
\label{appeq:gammaphi}
\een

{When the high frequency cutoff is below the relevant gate speeds, ${\w_h\ll2\pi/t}$,  the decay envelop Eq.~\eqref{Wlf} is  equivalent to the one computed from a Gaussian average over static noise with the variance given by the total integrated noise power, 
%related to the coefficient $c_\e$ by
\ben
{\s_\e=\<\de\e^2(t)\>=c_\e\sqrt{(1/\pi)\ln({\w_h}/{\w_l})}}\,,
\label{se}
\een
which can be significantly larger than $\s^*_\e$.  For example, for $\w_h=100$ GHz, $\s_\e=9~\mu$eV. 
However, as discussed in appendix \ref{sec:T1fit}, the noise spectrum is finite at GHz frequencies, so the limit ${\w_h\ll2\pi/t}$ is not satisfied. 

% For this \emph{static} noise approximation to be self consistent, the dephasing rate computed in this approximation, ${1/T_\phi^0=\s_\e/\sqrt{2}\hbar}$ must be faster than the high frequency cutoff, which requires  $\hbar\w_h\ll\s_\e$.  However, for the hybrid qubit, the short $T_1$ time near the charge qubit sweet spot, discussed in appendix \ref{T1fit}, indicate that the noise spectrum is finite at GHz frequencies, so this limit is not satisfied.}

%\mf{For fast gate speeds, not accurate to replace $\s'(t)$ with $\s_\e^*$
%However,  for $Z_\pi$ rotation with a faster gate period ${t_g=0.05}$ ns, $\s_\e'(t_g)=6.3$ that due to the time dependence in , using this variance to calculate a fidelity 
%This is a significant difference, sinceFor example,
% the infidelity calculated using  $\s_\e^*$ instead of  $\s'_\e(t_g)$ underestimates the infidelity by ${(\s^*_\e/\s'_\e(t_g))^2=84\%}$.}

\subsubsection{Dephasing due to high frequency noise\label{sec:hfnoise}}
\begin{figure}[t]
\begin{center}
\includegraphics[width=\linewidth]{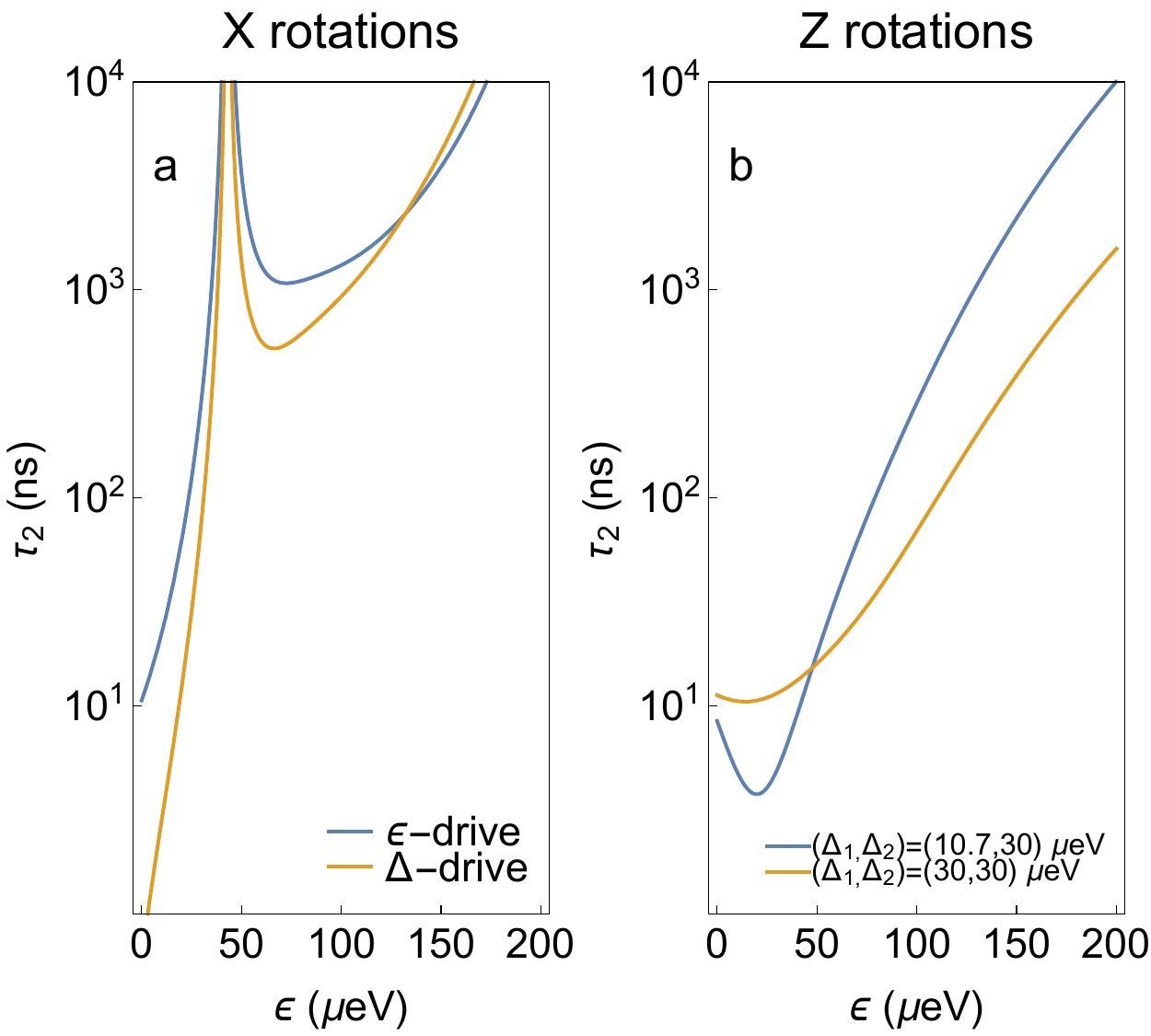}
\caption{(Color online) The exponential decay time scale $\tau_2$ as a function of detuning:
(a) For ac $X$ rotations,  with the tunnel couplings${(\D_1^0,\D_2^0,r)=(30,30)~\mu}$eV, for detuning ($\e$) and tunnel coupling ($\D$) drive.
(b) For dc $Z$ rotations,  with the tunnel couplings ${(\D_1^0,\D_2^0,r)=(30,30)~\mu}$eV used in this work and ${(\D^0_1,\D^0_2)=(10,30)~\mu}$eV used in Ref.~\onlinecite{kimCM15}.}
\label{fig:tau2}
\end{center}
\end{figure}
The high frequency component of the quadratic noise term $\de\e^2$  (Eq.~\eqref{gamma2}) causes an exponential decay $e^{-t/\tau_2}$, where $\tau_2$ is given in Eq.~\eqref{eq:tau2}, so that the total decay envelop in  \eq{W} is given by
\ben
W(t)=e^{-t/\tau_2}W_{\rm lf}(t)\,.
\label{W1}
\een
%\begin{figure}[t]
%\begin{center}
%\includegraphics[width=.5\linewidth]{Tau2FigDC}
%\caption{The exponential decay time scale $\tau_2$ as a function of detuning for DC gates in the \mf{dc part}}
%\label{Tau2FigDC}
%\end{center}
%\end{figure}
We plot the exponential decay time scale $\tau_2(X)$ for ac  $X$ rotations in Fig.~\ref{fig:tau2}.
This decay time is $>10~\mu$s near the optimal working point  ($\e>200\mu $eV), which, for the typical ac gate times considered in this work of $<10$ ns, causes infidelities of $<0.1\%$, and is thus negligible.  

The exponential decay time scale $\tau_2(Z)$ for dc  $Z$ rotations are plotted in Fig.~\ref{fig:tau2}, for both optimal tunnel couplings considered in this work and that of Ref.~\onlinecite{kimCM15}.  
Although this decay time  $\tau_2(Z)=1~\mu$s near $\e\sim200~\mu$eV is significantly shorter than $\tau_2(X)$ rotations, due to the short $Z$ gate periods $\sim 1$ ns,  $Z$ the gate infidelity, which can be estimated as $t_g(Z)/\tau_2(Z)\sim0.01\%$, is still very small, as noted in section \ref{sec:chargedisp}.
Note that this decay time is actually shorter for the optimal tunnel couplings in this work then that of Ref.\onlinecite{kimCM15}.  
This is because $\tau_2(Z)\propto G_{XX}\propto \w_R^2$, [cf. Eq.~\eqref{appeq:a12Z}] scales quadratically with the detuning driven Rabi frequency, which was increased in going from the tunnel coupling $\D_1=10~\mu$eV (Ref.~\onlinecite{kimCM15}) to $\D_1=30~\mu$eV (optimal) in this work.

%\begin{table}[b]
%\begin{center}
%{\begin{tabular}{lcc}
%\hline\hline
%&charge&hybrid\\
%\hline
%detuning $\e$~($\mu$eV)&20&~170\\
%{Rabi freq.} $f_R$~(GHz)&1-2&0.1-0.2\\
%{Excitation freq.} $f_Z$~(GHz)&4.5& 11.5\\
%AC X-rotation  $T_2$~(ns) &1.3&33\\
%DC Z-rotation  $T_2$~(ns) &1.3&10\\
%DC $X_{\pi/2}$ fidelty &85&--\\
%DC $Z_{\pi}$ fidelty   &95&96\\
%AC $X_{\pi}$ fidelty &85&93\\
%$T_1$~(ns) &7 &\\
%\hline\hline
%\end{tabular}}
%\end{center}
%\caption{Experimental values taken from \cite{kimNAT14,kimNATL15}
% with  $(E_{\rm ST},\D_1,\D_2)/h$ =$(50.2,10.8,30.6)~\mu$eV=$(12.1,2.6,7.4)$~GHz.}
%\label{values}
%\end{table}
\subsection{Simulations with numerically generated $1/f$ noise\label{sec:1overf}}
\begin{figure}[t]
\includegraphics[width=0.8\linewidth]{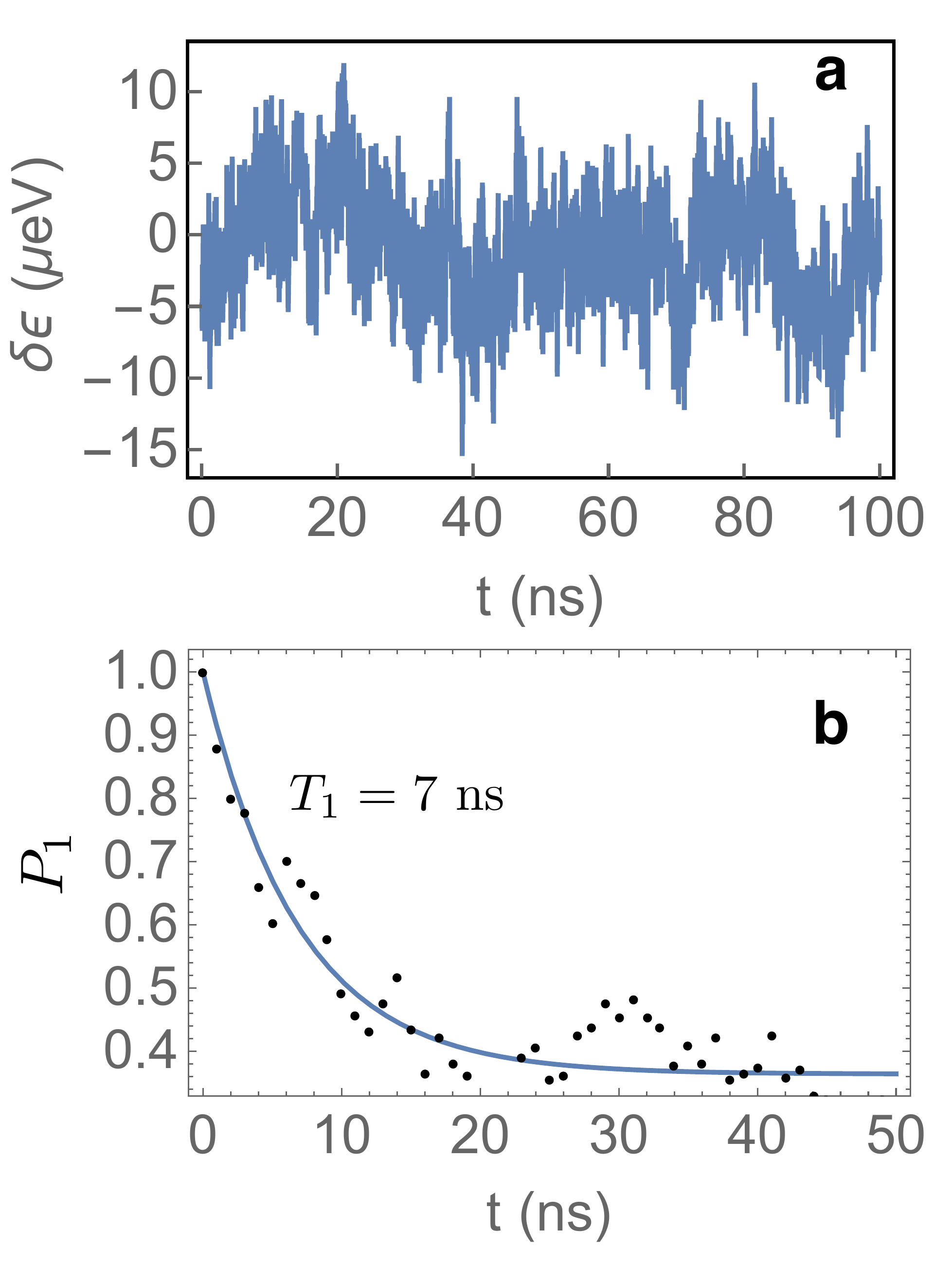}
\caption{(Color online) (a)  $1/f$ detuning noise generated numerically.
(b)  The probability of occupying $\ket{1}$ as a function of time (dots), computed by numerical integration of the master equation with detuning noise generated as shown in (a). (Solid line) Fit to an exponential decay with $T_1=7$ ns.}
\label{fig:T1fit}
\end{figure}
In this appendix, we describe our simulations of qubit dynamics using the three state Hamiltonian Eq.~\eqref{Hdc}, in the presence of numerically generated  $1/f$ detuning noise, following  the procedure described in \cite{press78}.
%In these simulations, we first check that, near the optimal parameter space for $(\e,A_{\e,g})$ and in the absence of noise, the leakage probability to $\ket{2}$ is negligible. 
%of order $10^{-4}$.
We first generate a white noise time series $f(t)$, whose Fourier component
\[f_\w=\int {dt'} e^{i\w t'}f(t')\]
is Gaussian distributed with a $\w$ independent variance, which we set equal to the detuning noise variance
\[S_f(\w)=\<|f_{\w}|^2\>=\s_f^2\,.\]
We then multiply the stochastic Fourier components by $1/\sqrt{\w}$,  and construct a  $1/f$ noise time series 
\ben
\de\e(t)=\int {d\w\over2\pi}e^{i\w t}\frac{f_{\w}}{\sqrt{\w}}\,,
\label{eq:F}
\een
that has a noise spectrum
\[S_\e(\w)=\left\langle\left|{f_{\w}\over\sqrt{\w}}\right|^2\right\rangle=\frac{\s_f^2}{\w}\,,\]
which gives the desired noise spectrum if we set white noise variance equal to the noise strength, $c_\e=\s_f$.  To check that this is the correct noise strength, we simulate $T_1$ relaxation by initiallizing the qubit in the state $\ket{1}$ and computing the probability to remain in $\ket{1}$ as a function time.  Fitting this probability to an exponential decay $P_1=e^{-t/T_1}+(1-e^{-t/T_1})P_{1f}$ yields the relaxation time  $T_1=7$ ns, consistent with the experimental result used to determine $c_\e$, see Fig.~\ref{fig:T1fit}. 

In our simulations, we find it convenient to use the discrete cosine transform, where all Fourier components are real from the outset (instead of the discrete Fourier transform). 
In computing the qubit fidelity,  the ideal gate is defined as the $X_\pi$ ac gate operation without any noise, see \eq{process}.
We find that the qubit fidelity  converge after averaging the solution to the density matrix equation of motion over 20 realization of the $\de\e(t)$ time series.  
% since ${\w_Z\gg\w_R}$ (for optimal parameters),  ``low frequency noise" in the RF is characterized by noise power at the qubit frequency, $S'_{XX}(\w\leq\w_R)\simeq S_{XX}(\w_Z)$.
%As a result
\bibliography{physics}
\end{document}